\documentclass[twocolumn]{aastex631}

\usepackage[utf8]{inputenc}
\usepackage{xcolor}
\usepackage{float}
\usepackage{placeins}

\begin{document}

\title{Discovery of Hidden Extragalactic Pulsars in the Parkes Archival Data}

\author[0000-0002-9409-3214]{Rahul Sengar}
\affiliation{Max Planck Institute for Gravitational Physics (Albert Einstein Institute), D-30167 Hannover, Germany \\}
\affiliation{Leibniz Universit\"{a}t Hannover, D-30167 Hannover, Germany\\}

\author[0000-0003-0292-4453]{L. Vleeschower}
\affiliation{Center for Gravitation, Cosmology, and Astrophysics, Department of Physics and Astronomy, University of Wisconsin-Milwaukee, P.O. Box 413, Milwaukee, WI 53201, USA\\}

\author[0000-0003-3294-3081]{M. Bailes}
\affiliation{Centre for Astrophysics and Supercomputing, Swinburne University of Technology, P.O. Box 218, Hawthorn, Victoria 3122, Australia\\}
\affiliation{Australian Research Council Centre of Excellence for Gravitational Wave Discovery (OzGrav)\\}

\author[0000-0002-2578-0360]{Fronefield Crawford}
\affiliation{Department of Physics and Astronomy, Franklin and Marshall College, P.O. Box 3003, Lancaster, PA 17604, USA\\}

\author[0000-0003-1301-966X]{Duncan R. Lorimer}
\affiliation{Department of Physics and Astronomy, West Virginia University,
PO Box 6315, Morgantown, WV 26506, USA\\}
\affiliation{Center for Gravitational Waves and Cosmology, West Virginia University, Chestnut Ridge Research Building, Morgantown, WV 26505, USA\\}

\author[0000-0002-3017-092X]{Joshua P. Ridley}
\affiliation{School of Engineering, Murray State University, 200 Sparks Hall, Murray, KY 42071, USA\\}

\

\begin{abstract}

Just over 1\% of the known radio pulsar population resides in the Large and Small Magellanic Clouds (MCs), the satellite galaxies that are 50--62\,kpc distant from the Milky Way. These pulsars are important for constraining the peak of the pulsar luminosity function, and may give insights into neutron star populations formed from stars of different metallicities than the Milky Way. Using search techniques that have proven highly effective at discovering a large number of pulsars in archival data by folding candidates near the noise floor at high harmonic folds, we reprocessed the original Parkes Multibeam MC surveys and discovered six pulsars. Five of these pulsars are located in the Large Magellanic Cloud (LMC), including one with a spin period of 3.8\,s, making it the slowest pulsar known in the MCs. The remaining pulsar was detected toward the Small Magellanic Cloud (SMC), and is associated with the Milky Way foreground. Four of the newly discovered pulsars exhibit scintillation, with one showing clear nulling and others showing possible nulling behavior. These discoveries bring the total number of radio pulsars known in the LMC to 49. A luminosity function and duty cycle analysis of all current MC pulsars shows that they are consistent with Galactic canonical pulsars. Population synthesis simulations suggest that deeper surveys would reveal many more pulsars in the MCs, including up to handful of millisecond pulsars, based on the adopted upper limits for their parent populations. Finally, this reprocessing demonstrates the continued importance of reanalyzing archival data, both to improve the completeness of pulsar surveys and detect scientifically valuable objects.

\end{abstract}


\section{Introduction} \label{sec:intro}

Although over 4300 pulsars have been discovered so far \footnote{See \url{https://www.atnf.csiro.au/research/pulsar/psrcat/}} \citep{manchester_05}, only about 1\% are extragalactic, located in nearby satellite galaxies called the Small and Large Magellanic Clouds (SMC and LMC). These Magellanic Clouds (MCs) are the only external galaxies for which a population of pulsars has been reported in the literature. Since the MCs are located at relatively well-constrained mean distances of about 50\,kpc for the LMC and 62\,kpc for the SMC \citep{pietrzy_19, graczyk_20}, their pulsars can be assigned characteristic distances with fractional uncertainties of order 6–11\% which are due to the line-of-sight depths of the Clouds \citep{hashcke_12, graczyk_20}. However, distances to most Galactic pulsars are inferred from dispersion measure (DM)  using electron density models. These estimates are strongly model and sightline dependent and can differ from parallax distances by factors of several for many pulsars \citep{Deller+2019, Price+2021}. Both MCs are relatively free from heavy foreground confusion, and therefore provide an ideal environment to study pulsar demographics, emission physics, and the connection between pulsar populations and star-forming regions beyond the Milky Way. 

The SMC hosts a particularly rich population of High-Mass X-ray Binaries (HMXBs), especially Be/X-ray binaries, with over a hundred confirmed and candidate systems cataloged \citep{haberl_16}. However, the higher metallicity of the LMC influences stellar mass-loss rates and accretion modes in HMXBs. As a result, the LMC produces fewer Be/X-ray binaries than the SMC, but a larger fraction of supergiant HMXBs and persistent X-ray sources \citep{antoniou_16}. Given these stellar environments and binary evolution channels, the MCs are expected to host a diverse population of binary pulsars, offering unique insights into neutron star evolution under different galactic conditions.

The first radio pulsar discovered outside our Galaxy was PSR~B0529$-$66 \citep{mcculloch_83} in the LMC, and then later PSR~B0540$-$69 was identified as a 50-ms Crab-like pulsar \citep{seward_84, Middleditch+1985, manchester_93, Marshall+2016}. A faster-spinning young X-ray pulsar, PSR~J0537$-$6910 with a spin period of $\sim 16$\,ms was found in the supernova remnant N157B of the LMC \citep{marshall_98}. The LMC also hosts the magnetar SGR~0526$-$66, located within the supernova remnant~N49 \citep{Kulkarni+2003}. The first radio pulsar population in the MCs was found with the Parkes/Murriyang radio telescope when \citet{mcconnell_91} conducted a 610-MHz survey and discovered four radio pulsars, three of which were located in the MCs (two in the LMC and one in the SMC), with one lying in the foreground. With the advent of the Parkes 20-cm multibeam (PM) receiver, survey speed improved significantly, owing to its increased sensitivity and larger field of view. Using this receiver, \citet{crawford_01} performed a survey of the SMC at $\sim$1.4\,GHz and found two new pulsars (one in the SMC and one in the foreground), along with an additional pulsar near the 30 Doradus region of the LMC. This region is one of the most active sites of massive star formation in the Local Group, making it a promising environment for the formation of young neutron stars and radio pulsars \citep[see e.g.,][]{schneider_18}.

A major breakthrough came when a full survey of the MCs was conducted using the Parkes 20-cm multibeam receiver at 1374\,MHz in 2000--2001 (hereafter, the PM survey) which substantially increased the known extragalactic pulsar population. The initial processing of these data resulted in the discovery of 14 new pulsars, of which 12 were associated with the MCs: three in the SMC and nine in the LMC \citep{manchester_06}. The limiting flux density of this survey was $\approx 0.12$\,mJy at 1.4\,GHz which enabled the first population studies of the MC pulsars. The dispersion measures (DMs) implied mean free-electron densities comparable to those in the Galactic disk, and the luminosity function was found to follow a power law with slope close to $-1$ which is consistent with that of the Galactic pulsar population. These findings support the conclusion that, aside from selection effects related to distance and survey sensitivity, the radio properties of pulsars in the MCs are broadly similar to those of pulsars in the Milky Way. 

Following the PM survey of the MCs, \citet{ridley_13} conducted a higher-resolution survey of these galaxies. Its sensitivity to slow pulsars was comparable to that of the PM survey, but its finer frequency and time resolution made it more sensitive to millisecond pulsars (MSPs). \citet{ridley_13} analyzed the high-resolution survey data and also reprocessed the archival PM survey using lower S/N thresholds, an extended DM range, and more intensive acceleration searches enabled by increased computing power. This reprocessing resulted in the discovery of five additional pulsars in the archival PM data and three more were discovered in the high-resolution survey. Using the then--current sample and a Bayesian framework, they constrained the pulsar population in the LMC beaming towards Earth to a broad but informative 95\% credible interval of $57{,}000^{+70{,}000}_{-30{,}000}$. For that survey configuration they adopted a limiting flux density of $\approx 0.05$\,mJy, corresponding to a 1.4-GHz pseudo-luminosity of $\approx 125$\,mJy\,kpc$^{2}$ at the distance of the LMC. They also found the LMC luminosity function to be consistent with that of Galactic ``normal'' pulsars and, based on the observed sample, suggested a maximum 1.4-GHz pseudo-luminosity of $\approx 1000$\,mJy\,kpc$^{2}$.

Complementary pulsar search techniques, such as identifying highly circularly polarized compact radio sources in continuum imaging surveys, are also helping to detect unusual pulsars that can be missed or are difficult to detect in traditional time-domain periodicity searches of large-scale pulsar surveys \citep{sengar_26a, sengar_26b}. These search methods are also contributing to the growing pulsar population of the MCs. For instance, the Australian Square Kilometre Array Pathfinder (ASKAP) Variables and Slow Transients (VAST) continuum survey \citep{murphy_13} identified a highly variable, strongly circularly polarized point source in the LMC. Targeted time-domain searches with MeerKAT and Parkes confirmed it as the 322.5-ms pulsar PSR~J0523$-$7125, with a high rotation measure of $+456$\,rad\,m$^{-2}$ \citep{wang_22}. In recent years, with the advent of highly sensitive radio interferometers, sensitive targeted surveys with MeerKAT have begun to significantly increase the pulsar population of the MCs. The TRAPUM\footnote{\url{http://www.trapum.org}} (TRAnsients and PUlsars with MeerKAT, \citealt{Stappers+Kramer2016}) Large Survey Project, targeting specific regions in both the SMC and LMC, covers a broader frequency range at L-band (856--1712\,MHz) than prior MC surveys conducted at Parkes. With $\sim2$-hr integrations per pointing, it is approximately three times more sensitive than those earlier surveys. This has already resulted in the discovery of seven new pulsars in the SMC (increasing the known SMC population by 50\%; \citealt{carli_24}) and nineteen in the LMC (increasing the known LMC pulsar population by $\sim$80\%; \citealt{prayag_24, prayag_26}). 

Apart from new discoveries by recent surveys, archival pulsar survey data remain immensely useful. With improved search algorithms and increased computational power, archival datasets continue to yield pulsars that were missed in first-pass processings. Through reprocessing of the PM survey \citep{manchester_01} and the High Time Resolution Universe (HTRU) pulsar survey of the southern Galactic plane \citep{keith_10}, \citeauthor{sengar_23} (\citeyear{sengar_23, sengar_25}) discovered over a hundred new pulsars. More recently, reprocessing efforts of the Parkes 70-cm pulsar survey \citep{manchester_96} and PM surveys of the MCs \citep{manchester_06} discovered two additional pulsars in the LMC \citep{hisano_22, xia_25}. 

Encouraged by these successes, and using similar search techniques to those employed by \citet{sengar_23} and \citet{sengar_25}, in this work we have reprocessed the archival data of the MC surveys conducted by \citet{manchester_06} and \citet{ridley_13}. The remainder of this paper is organized as follows. In Section~\ref{sec:data_reduction_and_searches}, we describe the data reduction and the FFT- and Fast Folding Algorithm (FFA)-based periodicity searches. In Section~\ref{sec:results}, we present the results, including the discovery of five pulsars in the LMC. In Section~\ref{sec:discussion}, we discuss the luminosities, duty-cycle properties of Galactic and MC pulsars as well as population synthesis results for the MCs. Finally, we conclude and summarize our main results in Section~\ref{sec:conclusion}.

\section{Data reduction and periodicity searches} 
\label{sec:data_reduction_and_searches}

The data used in this work were obtained from the Australia Telescope National Facility (ATNF) Data Access Portal \citep[DAP;][]{toomey_25} under the project IDs \texttt{P269} and \texttt{P743}. The \texttt{P269} dataset corresponds to the deep survey of the MCs conducted by \citet{manchester_06} with the Parkes 20-cm multibeam receiver (i.e., the PM survey). These observations were recorded in 1-bit format with an integration time of approximately 2.3\,hr (\(\sim\)8400\,s) per pointing, a sampling time of 1\,ms, and a bandwidth of 288\,MHz, split into 96 channels of 3\,MHz each, spanning roughly 1228--1516\,MHz. In total, this survey comprised 209 pointings (209$\times$13=2717 beams), 136 covering the LMC and 73 toward the SMC.

The \texttt{P743} dataset corresponds to the high-time-resolution observations described in \citet{ridley_13} (hereafter, the high-resolution survey). These data were also obtained with the Parkes 20-cm multibeam receiver and have a similar integration time per pointing, but a sampling time of 64\,\(\mu\)s and an effective bandwidth of 340\,MHz with 1024 channels (\(\sim\)390\,kHz each), and are therefore more sensitive to short-period pulsars. This dataset consisted of 74 pointings ($\sim$960 beams), which included 56 pointings targeting the LMC and 18 toward the SMC.

\subsection{FFT based periodicity searches}
\label{subsec:FFT_searches}

Prior to the main periodicity searches, the statistics of the frequency channels were computed using the \texttt{rfifind} suite from \textsc{PRESTO}\footnote{\url{https://github.com/scottransom/presto}} \citep{ransom_01, ransom_11}, after which the identified bad channels were removed from the raw filterbank files. The cleaned data were dedispersed and searched using an FFT-based pipeline built on the GPU-accelerated search code \textsc{PEASOUP}\footnote{\url{https://github.com/ewanbarr/peasoup}} \citep{sengar_23}, covering a DM range up to 500~pc\,cm$^{-3}$. We adopted this upper DM limit because no radio pulsar currently known in the MCs has a DM greater than $\sim 326$~pc\,cm$^{-3}$, with the highest-DM systems found in previous Parkes and recent MeerKAT searches remaining well below our search limit. 

Linear acceleration searches were carried out up to a maximum acceleration of $|a| = 25~\rm m\,s^{-2}$, which provides coherent sensitivity to binary pulsars with orbital periods of the order of one day, depending on orbital phase and companion mass. After the FFT stage, harmonic summing up to 32 harmonics  was carried out, and candidates with a spectral signal-to-noise ratio (S/N) $\geq 5.0$ were retained. On average, about 10,000 candidates per beam with spectral S/N $\geq 5.0$ were generated for the PM survey data and 16000 for the high resolution data. Since it is not feasible to fold such a large number of candidates per beam ($\sim 42$ million candidates in total), further filtering was necessary to select a manageable subset for folding and for the production of diagnostic plots used in visual inspection.  

Standard FFT-based searches are best suited for detecting rapidly-spinning ($P < 200$\,ms) pulsars, MSPs, and those with broad pulse profiles (duty cycle $\geq 10\%$) where the pulse train is well approximated by a sinusoid. However, more slowly-spinning ($P > 200$ ms) pulsars can have narrow duty cycles ($\leq 3\%$), consistent with the observed dependence of pulse width on spin period \citep[e.g.,][]{Maciesiak+2011, Johnston+2019}. Standard FFT searches face two key challenges when detecting such pulsars. The first is that the signal power of narrow pulses is spread across many harmonics in the FFT, and harmonic summing in many FFT searches is an incoherent process \citep{handbook_04}. Therefore, the resulting spectral S/N is sub-optimal. The second is the presence of red noise at the lower end of the frequency spectrum, which can either completely mask the pulsar signal or reduce its spectral S/N to the point where it becomes indistinguishable from noise candidates. The de-reddening algorithms employed by search codes are difficult to perfect, because the functional form of the red noise is not predictable and can change on many different timescales \citep{cameron_17, morello_20}.

To address these issues, we adopted the candidate selection criteria of \citet{sengar_23} and \citet{sengar_25}. Rather than folding all $\sim$10,000 candidates per beam, we selected a smaller subset of candidates to ensure coverage across the full ranges of spectral S/N, spin period ($P$), and harmonic sums ($nh$). The adopted selection rules are as follows:

\begingroup
\setlength{\arraycolsep}{2pt}
\renewcommand{\arraystretch}{1.05}
\[
\begin{array}{@{}l@{}l@{}}
\text{For } P<200~\mathrm{ms}: &
\left\{
\begin{array}{@{}l@{}}
S/N_{\mathrm{FFT}}\!\ge\!9.0,\ nh>1 \\[1pt]
S/N_{\mathrm{FFT}}\!\ge\!10.0,\ nh=1 \\[1pt]
8.0\!\le\!S/N_{\mathrm{FFT}}\!<\!9.0,\ 2<nh\!\le\!32 \\[1pt]
7.0\!\le\!S/N_{\mathrm{FFT}}<8.0,\ nh\!\ge\!8
\end{array}
\right.
\end{array}
\]
\endgroup

\begingroup
\setlength{\arraycolsep}{2pt}
\renewcommand{\arraystretch}{1.05}
\[
\begin{array}{@{}l@{}l@{}}
\text{For } P>200~\mathrm{ms}: &
\left\{
\begin{array}{@{}l@{}}
S/N_{\mathrm{FFT}}\!\ge\!7.5,\ nh>1,\\[1pt]
S/N_{\mathrm{FFT}}\!\ge\!10.0,\ nh=1 \\[1pt]
7.0\!\le\!S/N_{\mathrm{FFT}}\!<\!7.5,\ 2<nh\!\le\!16\\[1pt]
5.5\!\le\!S/N_{\mathrm{FFT}}<7.0,\ nh\!\ge\!8,\ P>500~\mathrm{ms}. \\[1pt]

\end{array}
\right.
\end{array}
\]
\endgroup

\begin{figure*}
\centering
\includegraphics[width=2.2\columnwidth]{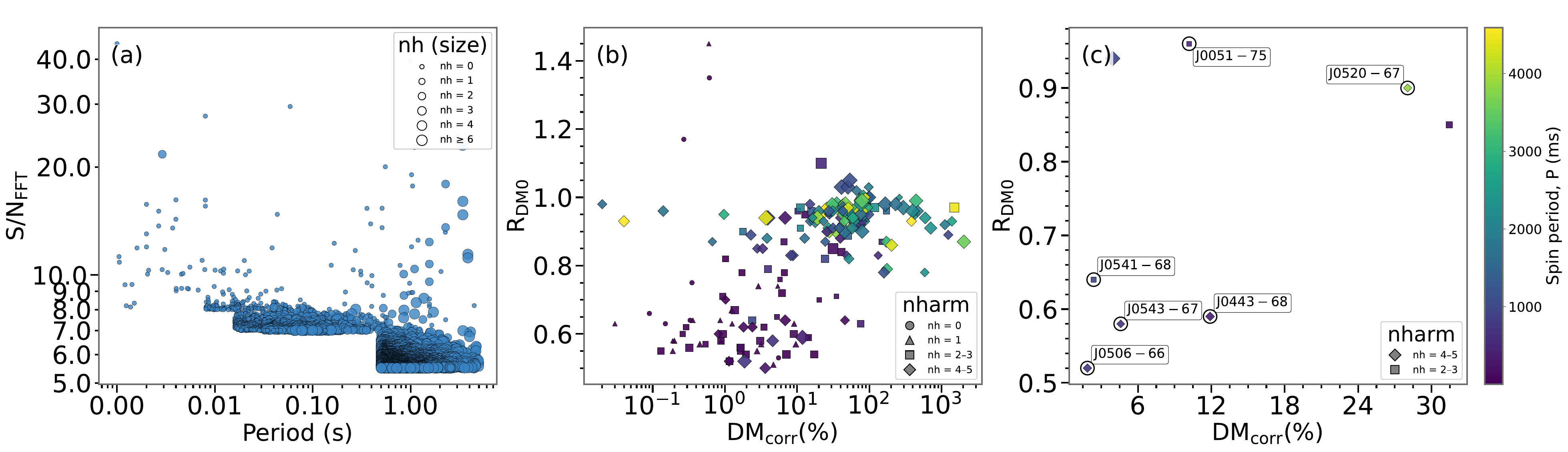}
\caption{Plots illustrate an example of the candidate selection and filtering procedure used to identify pulsars for visual inspection for six beams in which pulsars were found. \textit{Left:} Distribution of candidates after applying the selection rules to the raw candidate files from beams in which new pulsars were found. The legend shows \texttt{nharm}, the harmonic–sum index in the FFT search (\texttt{nharm} = 0,1,2,3,4,5), where index $n$ corresponds to $2^n$ harmonics (i.e., 1, 2, 4, 8, 16, 32). 
\textit{Middle:} R$_{\rm DM0}$ as a function of DM correction for candidates that survived folding, with marker color indicating the spin period range and marker shape denoting the \texttt{nharm} range. \textit{Right:} 
Final set of candidates selected for visual inspection, in which only the pulsars and two spurious candidates survived the filtering criteria.}
\label{fig:cand_filtering}
\end{figure*}

Using the above criteria, the average number of candidates selected for folding per beam was approximately 200 (see Figure~\ref{fig:cand_filtering}(a)). The folding procedure was similar to that described by \citet{sengar_23}. In brief, we used \textsc{dspsr}\footnote{\url{http://dspsr.sourceforge.net}}  \citep{vanStraten_11} to create archive files, which were subsequently cleaned using \texttt{clfd}\footnote{\url{https://github.com/v-morello/clfd}} \citep{morello_19}. The cleaned archives were then processed with the \texttt{pdmp} module from the \textsc{PSRCHIVE} software package\footnote{\url{https://psrchive.sourceforge.net/}} \citep{hotan_2004}, which provided both diagnostic plots and optimized folding parameters and S/Ns. Candidates with a folded S/N $<8.0$ or a DM $<2$\,pc\,cm$^{-3}$ were rejected. The surviving candidates were re-folded using the optimized values of period, DM, and acceleration, and any further candidates with DM $<2$ or folded S/N $<8$ were discarded. An example of these surviving candidates is shown in Figure~\ref{fig:cand_filtering}(b) where low-S/N and near-zero-DM candidates have already been removed. However, a large number of candidates with a wide range of periods and DMs still remain, including radio frequency interference (RFI) candidates. Therefore, to further reduce the number of unwanted candidates we also folded each candidate at zero DM using the \textsc{PSRCHIVE}'s tool \texttt{pam} to compute the ratio of the folded S/N at $\rm DM=0$ to that at the optimized DM (DM$_{\rm opt}$), and defined as

\[
R_{\rm DM0} =
\frac{({\rm S/N})_{\rm DM=0}}
{({\rm S/N})_{\rm opt}}
\]

The rationale behind this criterion is that genuine dispersed pulsar signals are expected to show an improvement in the folded S/N at the DM$_{\rm opt}$ compared to $\mathrm{DM}=0$, whereas RFI typically does not. Therefore, for real pulsars, $R_{\rm DM0}$ would be less than unity. In addition, we derived other diagnostic parameters, including the duty cycle of each candidate and the DM correction, DM$_{\rm corr}$, which measures how much the DM$_{\rm opt}$ differs from the DM obtained during the FFT search (DM$_{\rm FFT}$):

\[
\mathrm{DM}_{\mathrm{corr}} =
\frac{\left| \mathrm{DM}_{\mathrm{opt}} - \mathrm{DM}_{\mathrm{FFT}} \right|}
{\mathrm{DM}_{\mathrm{FFT}}}
\]

and the DM error fraction, DM$_{\rm err\,frac}$, which is defined as the ratio of the uncertainty on the optimized DM ($\sigma_{\mathrm{DM_{opt}}}$) to the optimized DM itself,
\[
\mathrm{DM_{err\;frac}} = \frac{\sigma_{\mathrm{DM_{opt}}}}{\mathrm{DM_{opt}}}.
\]

The reason behind using $\mathrm{DM}_{\mathrm{corr}}$ and $\mathrm{DM_{err\;frac}}$ parameters is that a genuine pulsar DM would require only modest corrections to the initially detected DM in the FFT search after folding optimization and would show maximum S/N as a function of DM, resulting in relatively small uncertainties in the DM$_{\rm opt}$. In contrast, spurious candidates or RFI often produce poorly constrained or unstable DM solutions which lead to larger values of DM$_{\rm corr}$ and DM$_{\rm err\,frac}$.

From the known pulsar detections in the survey, we empirically identified suitable limits for these parameters that efficiently identify all known pulsars while rejecting a large fraction of false positives. These parameters were used to filter candidates prior to visual inspection. Specifically, candidates were selected if they satisfied the following criteria:

\[
\mathrm{R_{DM0}} < 1,\quad 
\mathrm{DM_{corr}} < 35\%,\quad 
\mathrm{DM_{err\;frac}} < 35\%.
\]

The full SMC and LMC survey data yielded \(\sim\)3,500 candidates with $P > 200\,\mathrm{ms}$ and \(\sim\)8,000 candidates with $P < 200\,\mathrm{ms}$ requiring visual inspection (see Figure~\ref{fig:cand_filtering}(c)).

\begin{deluxetable*}{lcccccccc}
\tablecaption{Properties of newly discovered pulsars. The pulsar name (JName), right ascension (RA), and declination (Dec) correspond to the discovery designation, while the Obs\_ID denotes the observation/beam in which the pulsar was discovered in the MC survey data. The positional error in right ascension and declination is estimated either from the beam size of the MB receiver or redetection of the pulsar in other beams. The spin period, DM, their uncertainties, and folded S/N were measured from the discovery observation using the PSRCHIVE \texttt{pdmp} routine, which searches over period and DM to maximize the folded S/N. The parameter $w_{50}$ represents the full width at half-maximum (FWHM) of a Gaussian fit to the discovery pulse profile. The listed flux densities at 1.4\,GHz ($S_{1400}$) represent lower-limit estimates derived from the radiometer equation. }
\label{tab:new_pulsars}
\tablehead{
\colhead{JName} & 
\colhead{ObsID} & 
\colhead{RA} & 
\colhead{Dec} & 
\colhead{Period} & 
\colhead{DM} & 
\colhead{S/N} & 
\colhead{$w_{50}$} & 
\colhead{$S_{1400}$} \\
 & & (J2000) &  (J2000) & (ms) & (pc cm$^{-3}$) & & (ms) & ($\mu$Jy)
}
\startdata
J0051$-$75\tablenotemark{\scriptsize a} & SMC020\_004B1 & 00:51(2) & $-$75:50(7) & 552.6966(3) & 25(8) & 22.4 & 56 & 150 \\
J0443$-$68\tablenotemark{\scriptsize b} & SMC013\_03241 & 04:43(1) & $-$68:01(7) & 366.9758(8) & 61.3(0.4) & 10.7 & 4.3 & 26.0 \\
J0506$-$66\tablenotemark{\scriptsize b,c}  & PLMC080\_04 & 05:06(1) & $-$66:57(7) & 740.8268(5) & 62(4) & 11.4 & 18.0 & 34.6 \\
J0520$-$67\tablenotemark{\scriptsize d} & SMC017\_03921 & 05:20(2) & $-67$:47(14) & 3852.65(2) & 151(50) & 19.0 & 64.9 & 44.8 \\
J0541$-$68\tablenotemark{\scriptsize b} & SMC019\_02261 & 05:41(1) & $-68$:29(7) & 953.166(4)  & 171(20) & 11.7 & 59.3 & 60.3 \\
J0543$-$67\tablenotemark{\scriptsize e} & SMC019\_01921 & 05:43(1) & $-67$:37(7) & 648.359(2)  & 70(4)   & 13.6 & 8.4  & 26.2 \\
\enddata
\tablenotetext{a}{Scintillating pulsar,} independent discovery. Initially detected in the SMART survey.\vspace{-0.6em}
\tablenotetext{b}{Nulling pulsar,} confirmed using Murriyang's UWL receiver.\vspace{-0.6em}
\tablenotetext{c}{Detected in the high-res survey data of MCs. \vspace{-0.6em}}
\tablenotetext{d}{Confirmed in two adjacent observations (SMC019\_01671 and SMC017\_02271) at different epochs.\vspace{-0.6em}}
\tablenotetext{e}{Shows possible} variability in time and frequency, with possible nulling.

\end{deluxetable*}

\begin{figure*}
\centering
  \includegraphics[width=1.8\columnwidth]{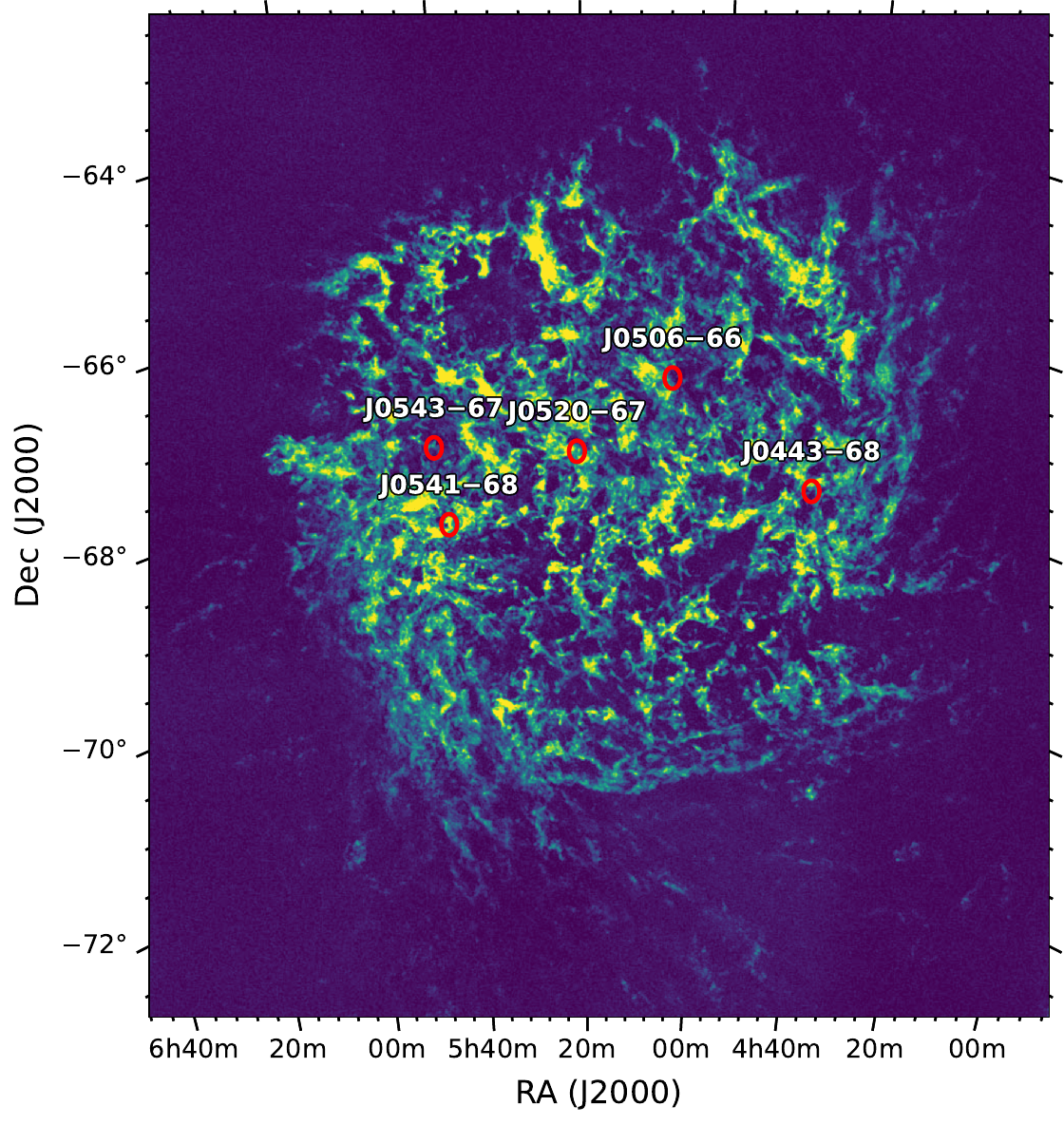}
  \caption{HI image of the Large Magellanic Cloud adapted from \citet{kim_98}. The red circles mark the positions of the newly discovered pulsars PSRs J0443$-$68, J0506$-$66, J0520$-$67, J0541$-$68, and J0543$-$67.}
  \label{fig:HI_plot}
\end{figure*}

\subsection{FFA-based periodicity searches}

In addition to the FFT-based searches, we performed time-domain periodicity searches using the Fast Folding Algorithm as implemented in the FFA-based \textsc{riptide}\footnote{\url{https://github.com/v-morello/riptide}} code \citep{morello_20}. As discussed in section \ref{subsec:FFT_searches}, FFT-based searches are less sensitive to narrow-duty-cycle and slow pulsars. However, by operating directly in the time domain, the FFA is less affected by red noise and retains sensitivity to narrow, non-sinusoidal pulse profiles by coherently folding the full time series at many trial periods. This improves sensitivity to long-period, narrow-duty-cycle pulsars compared to standard FFT-based searches.

\begin{figure}
\centering
  \includegraphics[width=0.975\columnwidth]{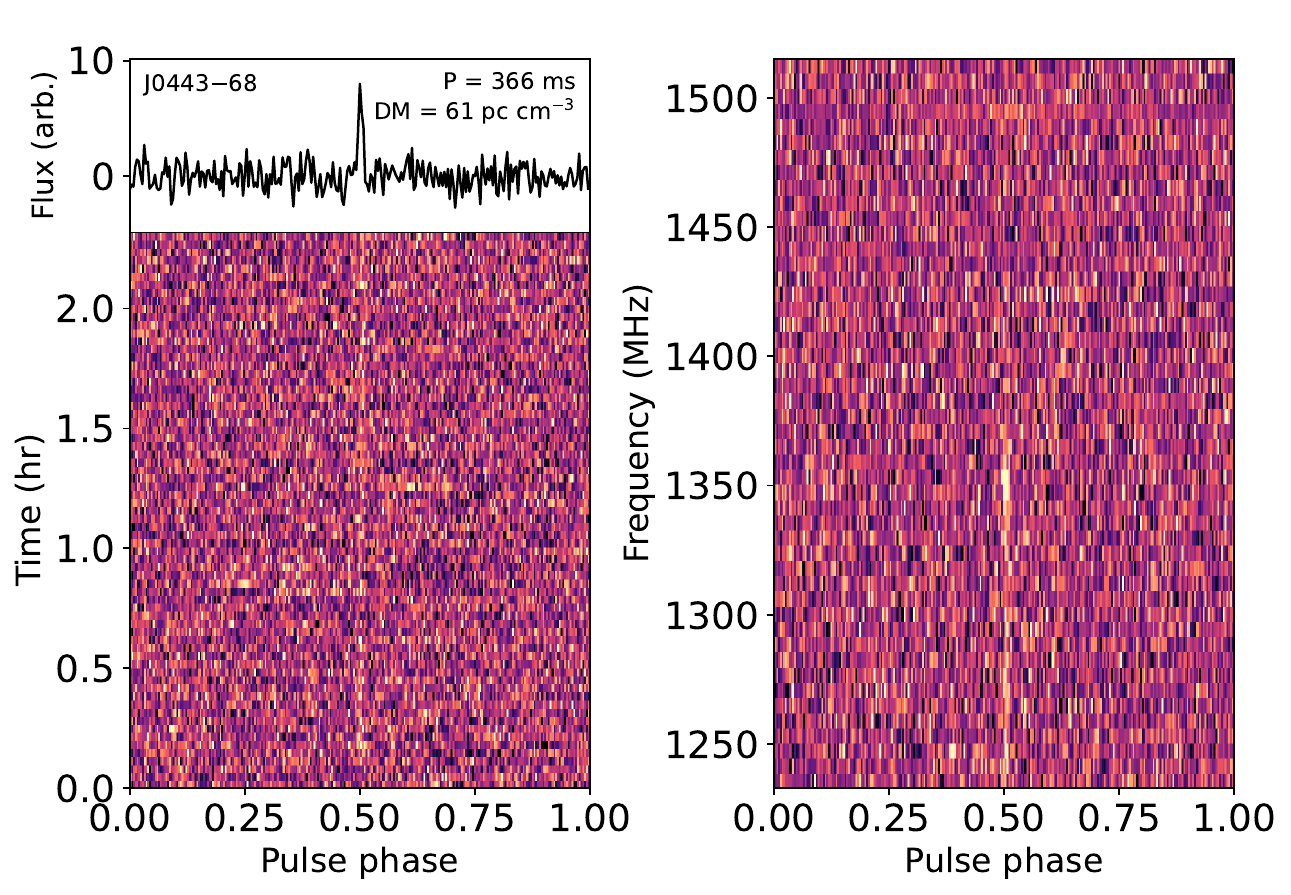}
  \includegraphics[width=0.975\columnwidth]{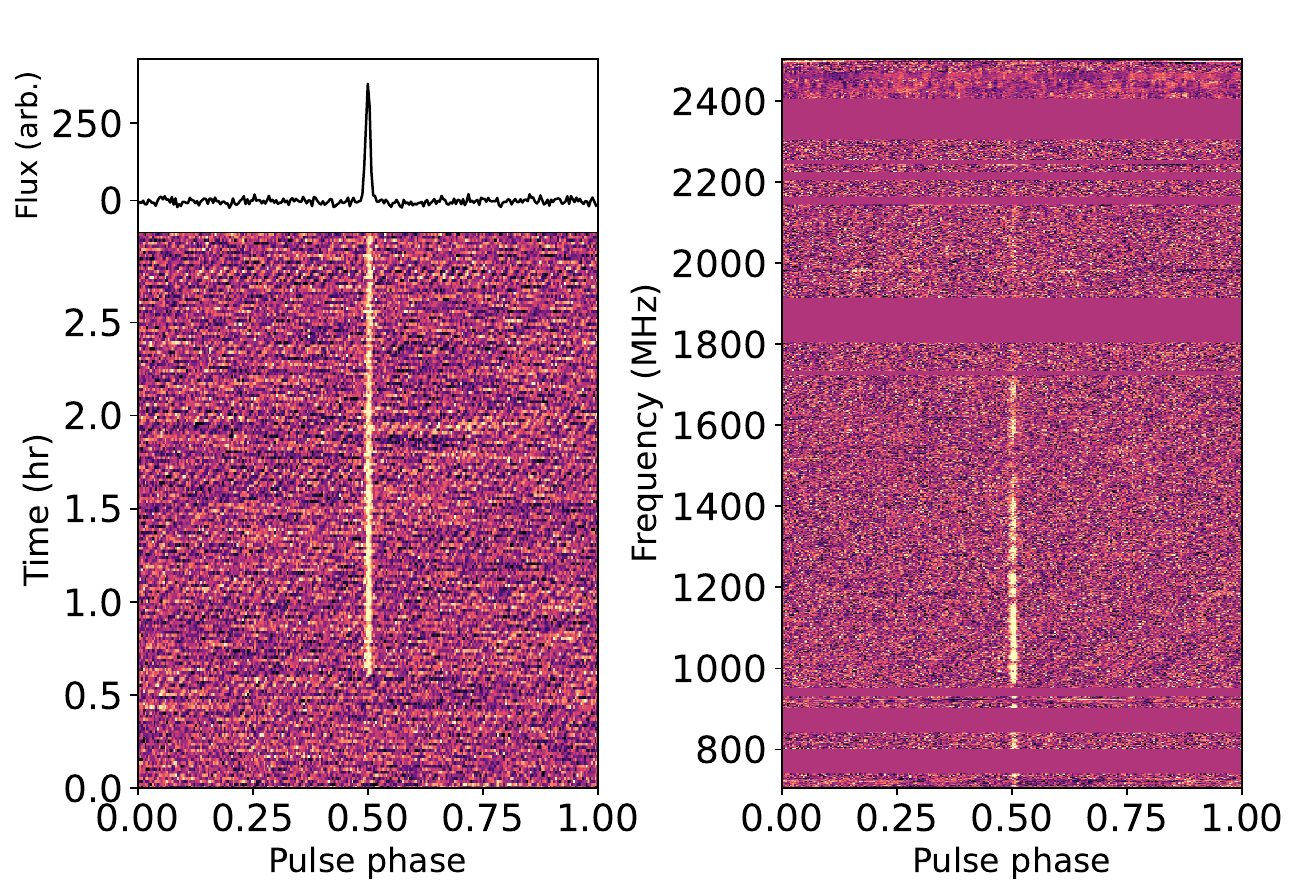}
  \caption{Discovery and confirmation plots of PSR~J0443$-$68. Top row: Discovery observation showing the time--phase plot (left) along with the integrated pulse profile (top subpanel) and the frequency--phase diagnostic plot (right). Bottom row: Confirmation observation obtained using the Parkes UWL receiver over 704-2500\,MHz, showing the corresponding time--phase and frequency--phase diagnostic plots.}
  \label{fig:J0443-68_plots}
\end{figure}

\begin{figure}
\centering
  \includegraphics[width=0.99\columnwidth]{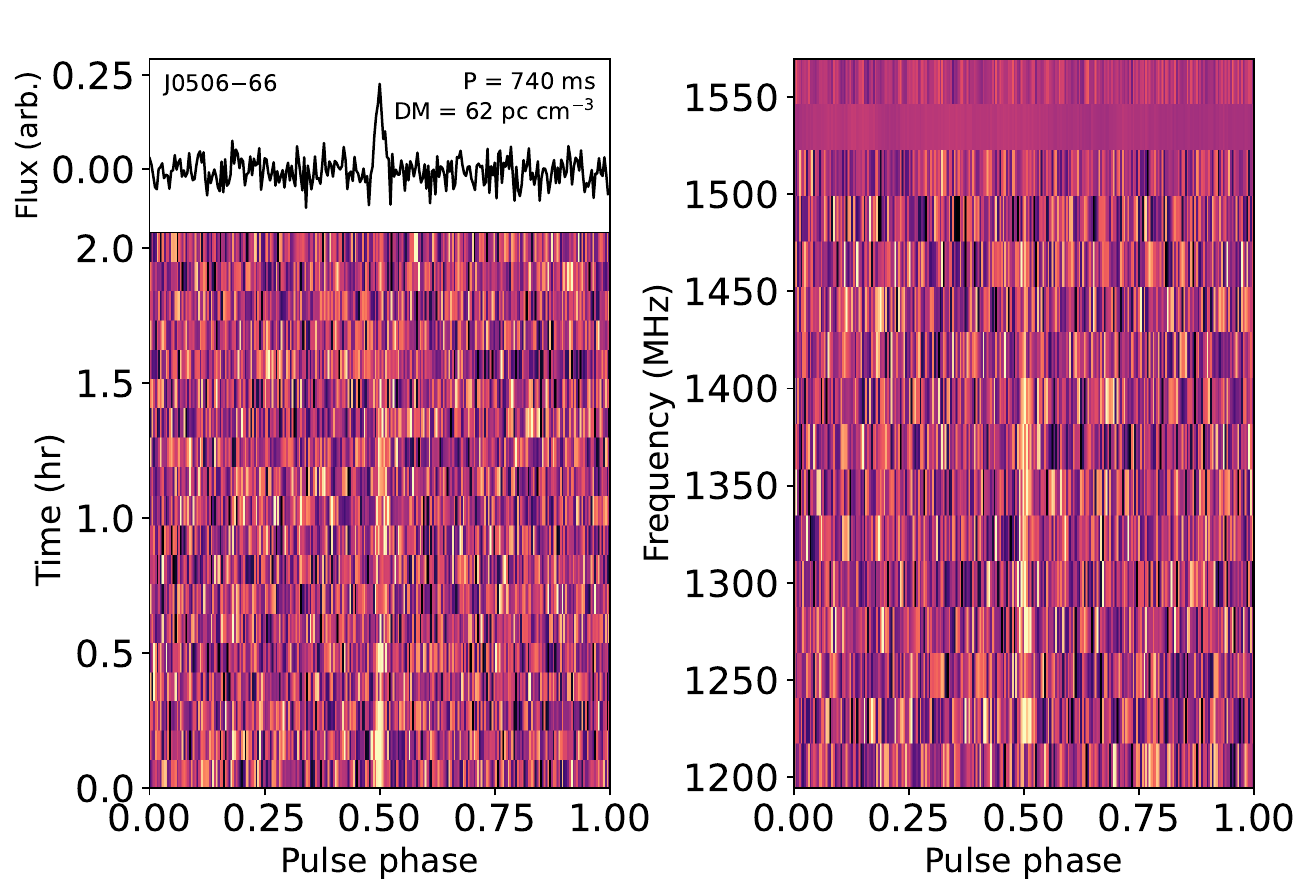}
  \includegraphics[width=0.99\columnwidth]{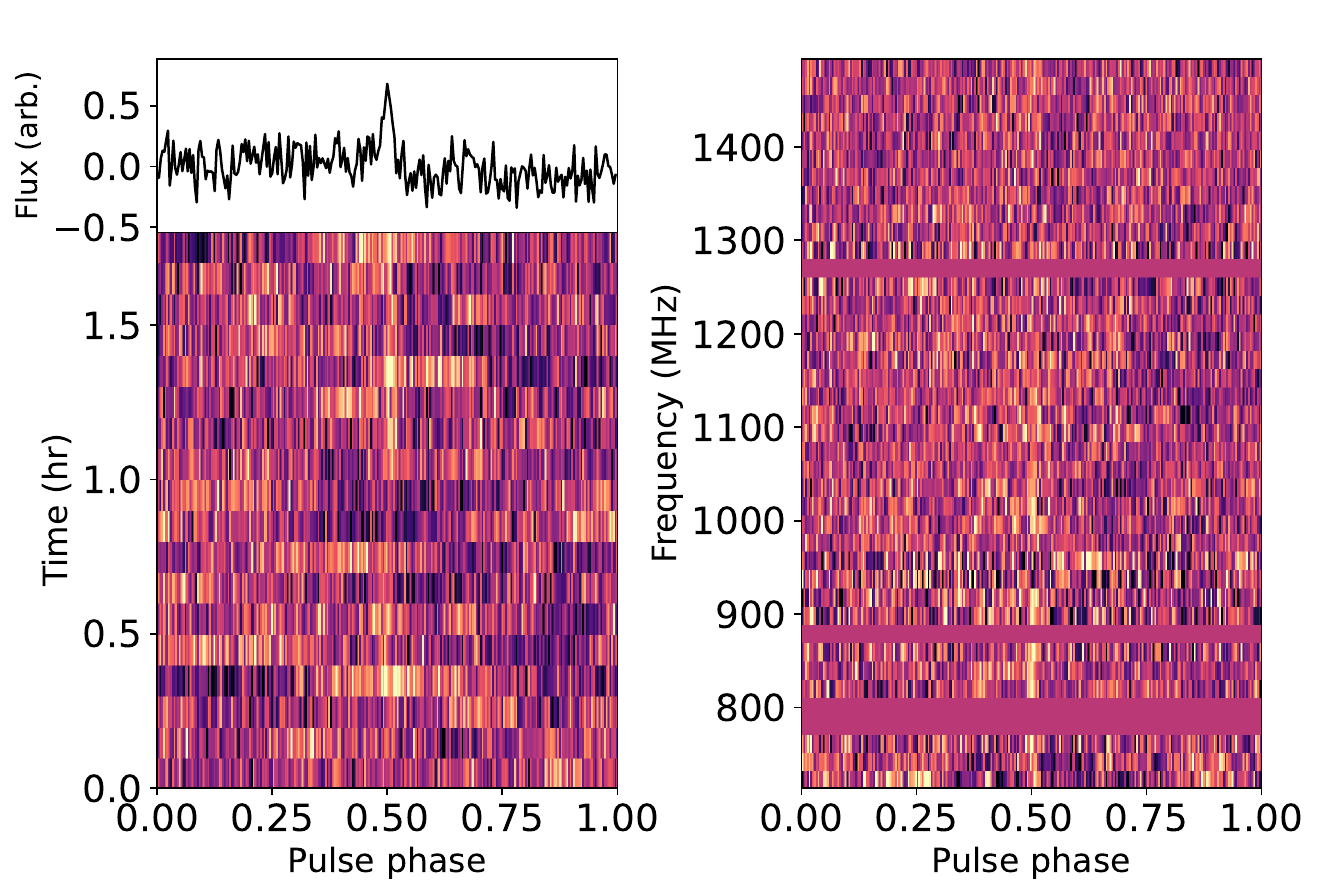}
  \caption{Discovery and confirmation plots of PSR~J0506$-$66. Top row: Discovery observation showing the time--phase plot (left) along with the integrated pulse profile (top subpanel) and the frequency--phase diagnostic plot (right). Bottom row: Confirmation observation obtained using the Parkes UWL receiver over 704-1500\,MHz, showing the corresponding time--phase and frequency--phase diagnostic plots.}
  \label{fig:J0506-66_plots}
\end{figure}

Following a similar initial procedure to that used in the FFT searches, \texttt{rfifind} was first used to identify and remove strong narrow-band and transient RFI from each pointing. Dedispersed time series were then produced over the same DM range as used in the FFT searches using \texttt{prepsubband} from \textsc{PRESTO}. The FFA stage coherently folded each time series over a dense grid of trial periods spanning 200~ms to 30~s. Candidates with an FFA S/N $\geq 8.0$ were retained for follow-up, folded using \textsc{dspsr}, and inspected using diagnostic plots produced with \textsc{PSRCHIVE}'s \texttt{pdmp}. No acceleration trials were applied to the time-series data, hence the FFA search was sensitive primarily to isolated pulsars.


\section{Results}
\label{sec:results}

\begin{figure*}
\centering
  \includegraphics[width=1.99\columnwidth]{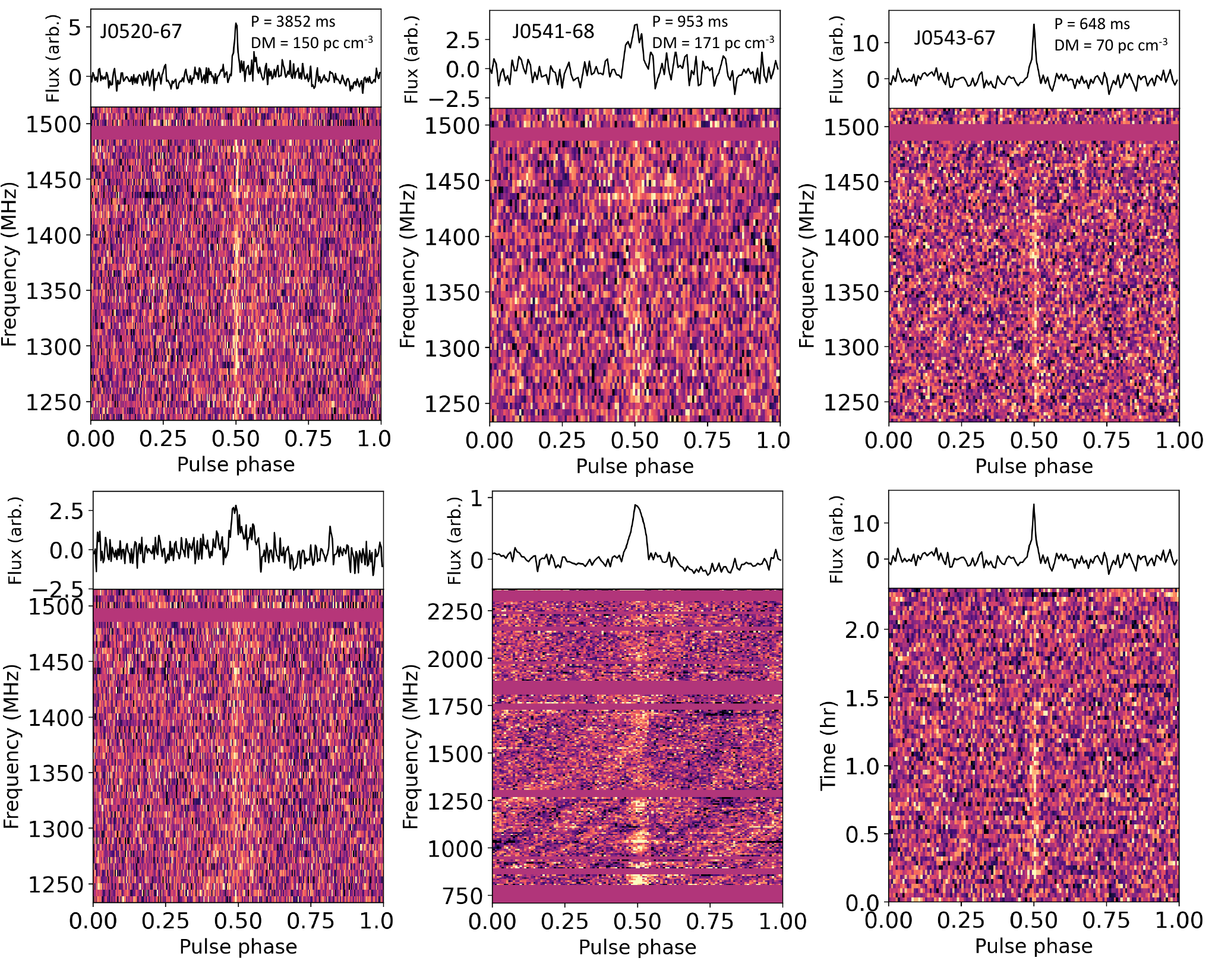}
\caption{Discovery and confirmation plots of three new pulsars in the Large Magellanic Cloud. Top row: Discovery observations of PSRs~J0520$-$67, J0541$-$68, and J0543$-$67, with the integrated pulse profile (top subpanel) and frequency--phase diagnostic plot (bottom subpanel). Bottom row: Confirmation of PSR~J0520$-$67 in a neighboring beam (left) and J0541$-$68 with the Parkes UWL receiver spanning 704--4032\,MHz (middle). Since PSR~J0543$-$67 was not redetected, we instead show its time-phase plot as further evidence of its authenticity (right).}
\label{fig:discovery_plots}
\end{figure*}

From the reprocessing of both the PM survey and the high-resolution survey of the MCs, we have discovered six new pulsars, five of which are located in the LMC. Their positions are shown in Figure~\ref{fig:HI_plot} and their properties are given in Table~\ref{tab:new_pulsars}. The positional uncertainties are typically $\sim7'$ for sources localized within a single Parkes multibeam beam, with larger uncertainties for sources constrained from detections in multiple adjacent beams. Since, recent MeerKAT/TRAPUM searches of the LMC were targeted, most of the new LMC pulsars reported in this work lie outside, or close to the half-power boundary of the nearest MeerKAT primary-beam pointings. The exception is PSR~J0520$-$67, whose position lies within one of the targeted fields. However, whether its position was covered by the coherent tied-array beam tiling requires a more detailed comparison with the survey beam footprint. All six pulsars were detected through our FFT-based periodicity searches. Five of them were discovered in the PM survey data, while one was found in the high-resolution survey data. We also redetected all six pulsars in the FFA-based search results, although no additional compelling candidates were identified. None of these pulsars show any evidence of acceleration either in their discovery observations or in the confirmation observations, suggesting that they are likely isolated systems. Since timing solutions are not yet available, we briefly discuss their observed properties below.

\subsection{PSR~J0443--68}
\label{subsec:J0443}

PSR~J0443$-$68 has a spin period of $P=366~\mathrm{ms}$ and $\mathrm{DM}=61~\mathrm{pc\,cm^{-3}}$. It was initially detected as a faint candidate with a folded S/N of 10.7. In its discovery observation, the pulsar emission was confined to approximately half of the observing frequency band ($1228$--$1372~\mathrm{MHz}$). An initial $1.5$\,hr follow-up with Parkes failed to redetect this pulsar. However, a subsequent $3$\,hr observation confirmed the source with a very high S/N. Using the Ultra-wideband low-frequency \citep[UWL;][]{hobbs_20} receiver, the pulsar was detected across much of the $700$--$1750~\mathrm{MHz}$ band with an $\mathrm{S/N}=70$ and some hint of the signal from 1750--2200 MHz (see bottom row of Figure~\ref{fig:J0443-68_plots}). Considering only the PM survey band (1228--1516\,MHz), the pulsar has an $\mathrm{S/N}=31$ which is $\sim 3$ times its discovery S/N. The frequency-pulse-phase plots from both the discovery and Parkes confirmation observations (see Figure~\ref{fig:J0443-68_plots}) show that the pulsar appears to switch on and off across the 1150--1700\,MHz band which indicates that the pulsar is scintillating. Using this observation, we attempted to measure the scintillation properties of the pulsar via the auto-correlation function (ACF) of the dynamic spectrum. The dynamic spectrum was formed from the on-pulse emission after subtracting the off-pulse baseline and masking the RFI in the affected channels and subintegrations. The 2D ACF was computed using a Fourier-domain correlation method, and Gaussian functions were fitted to the frequency and time slices of the ACF to estimate the scintillation bandwidth, $\Delta \nu_{d}$, and scintillation timescale, $\tau_d$. 

The ACF is dominated by a narrow zero-lag correlation spike centered at $(\Delta t,\Delta \nu)=(0,0)$. This feature is confined to approximately one subintegration and only a few frequency channels, indicating that the scintillation pattern is only marginally resolved at the available time and frequency resolution.

A formal Gaussian fit to the frequency-domain ACF yields $\Delta \nu_{d} \approx 1.6$\,MHz, which is within the range of the previously reported values for the MC pulsars \citep{johnston_21}. However, the corresponding time-domain fit yields $\tau_d$ smaller than the subintegration duration, demonstrating that the scintillation timescale cannot be reliably measured in this observation. Furthermore, after excluding the zero-lag component, the remaining ACF structure is weak and poorly constrained, suggesting that the measured $\Delta \nu_{d}$ should also be treated with caution.

Using the pulse profile from the discovery observation, the flux density is $\sim 26~\mu\mathrm{Jy}$. However, in its confirmation observation, its flux density increases to $\sim 75~\mu\mathrm{Jy}$, indicating that it is a highly variable source. During the confirmation observation, the pulsar enters a null state lasting $\sim 36$\,min. The pulsar was also not detected during the entire initial $1.5$\,hr follow-up observation. This suggests that the null state can persist for at least $\sim1.5$\,hr. The discovery of this pulsar is one of the classic examples of the effectiveness of the search techniques described in Section~\ref{subsec:FFT_searches} for uncovering such faint pulsars. The pulsar was found very near our adopted FFT noise floor with a spectral $\mathrm{S/N}$ of only $5.8$, and only when harmonic summing up to the 32nd harmonic was applied. It might not have been detectable above the noise at all if traditional harmonic summing limited to 16 harmonics had been used. Historically, the majority of pulsar search pipelines did not fold candidates of such low significance, with typical thresholds of 5-8. Interestingly, we also found this pulsar in our FFA-based search results, as expected, but with an S/N of only 8.5 which is about 20\% less than the S/N we obtained from our folding pipeline. The difference in S/N is not surprising as it could likely be due to differences in the way S/N is measured using different pipelines. The finite phase resolution, matched-filter width, possible downsampling of the narrow pulse profile, and small mismatches in the trial period and DM can contribute to some loss in S/N.

\subsection{PSR~J0506--66}

PSR~J0506$-$66 is a $740~\mathrm{ms}$ pulsar and has a DM of 62~pc\,cm$^{-3}$. We found this pulsar in the high-resolution LMC data as a faint candidate with a folded $\mathrm{S/N}=11.4$, and it was subsequently confirmed with a comparable S/N in a $\sim 1.8$\,hr Parkes UWL follow-up observation. In the discovery observation, detectable emission is largely confined to $\sim 1125$--$1400~\mathrm{MHz}$, whereas in the Parkes UWL confirmation it is predominantly visible only between $\sim 700$ and $900~\mathrm{MHz}$ (see Figure~\ref{fig:J0506-66_plots}). This frequency-dependent variability could be caused by scintillation, although the relatively low S/N of the detections prevents us from establishing scintillation as its origin. In its discovery observation, the longest interval over which the emission is undetectable is $\sim30$\,min which may indicate that similar to PSR~J0443$-$68, this pulsar also exhibits nulling behavior. However, the available S/N is insufficient to distinguish intrinsic nulling from scintillation or a low-flux state. The available S/N is also insufficient to reliably measure the scintillation bandwidth. We therefore interpret the observed emission as possibly showing both time- and frequency-dependent variability, with scintillation and possible nulling as plausible explanations. In our FFT searches, PSR~J0506$-$66 was detected with 32-harmonic summing and a spectral $\mathrm{S/N}$ of 6.5, suggesting that it would likely have been missed under a more conservative 16-harmonic summing regime and/or a higher spectral S/N cut.
 
\subsection{PSR~J0520--67}

PSR~J0520$-$67 is notable for being the slowest spinning radio pulsar known in either the SMC or LMC with a spin period of $P=3.8$\,s and a $\mathrm{DM}=150~\mathrm{pc\,cm^{-3}}$. It was first detected with a folded S/N of 19. During candidate inspection, the pulsar was independently confirmed in a neighboring beam of a different pointing which was observed 93\,days later and located $14'$ from the highest-S/N detection. To assess whether the source was present in any additional adjacent beams, we folded the data from all nearby pointings at the discovery spin period and DM, resulting in a tentative detection with $\mathrm{S/N}=8$ in another beam, also located about $14'$ from the primary discovery position (i.e., the beam yielding the highest folded S/N). We subsequently observed this pulsar at the coordinates corresponding to its highest-S/N detection for 2~hr with Murriyang and using the UWL receiver, but did not redetect it. In addition, we observed it at the second-highest detection position, as well as the intermediate position between all three beams, but did not detect it there either. Both the discovery and the confirmation plots for this pulsar are shown in the left panel of Figure~\ref{fig:discovery_plots}. Dedicated gridding observations are required to refine the position and redetect this pulsar.

\subsection{PSR~J0541--68}

PSR~J0541$-$68 has a spin period of $953~\mathrm{ms}$ and a $\mathrm{DM}=170~\mathrm{pc\,cm^{-3}}$. It was initially detected with a $\mathrm{S/N}=11.7$ and an estimated duty cycle of $\sim 6\%$. We subsequently observed this pulsar for $\sim 2$\,hr with the UWL receiver on the Parkes radio telescope and confirmed it with high significance ($\mathrm{S/N}=32$). As illustrated in the middle panel of Figure~\ref{fig:discovery_plots}, the UWL observation shows that the emission is detectable across the band from $\sim700~\mathrm{MHz}$ to $2000~\mathrm{MHz}$.

\begin{figure}
\centering
  \includegraphics[width=0.99\columnwidth]{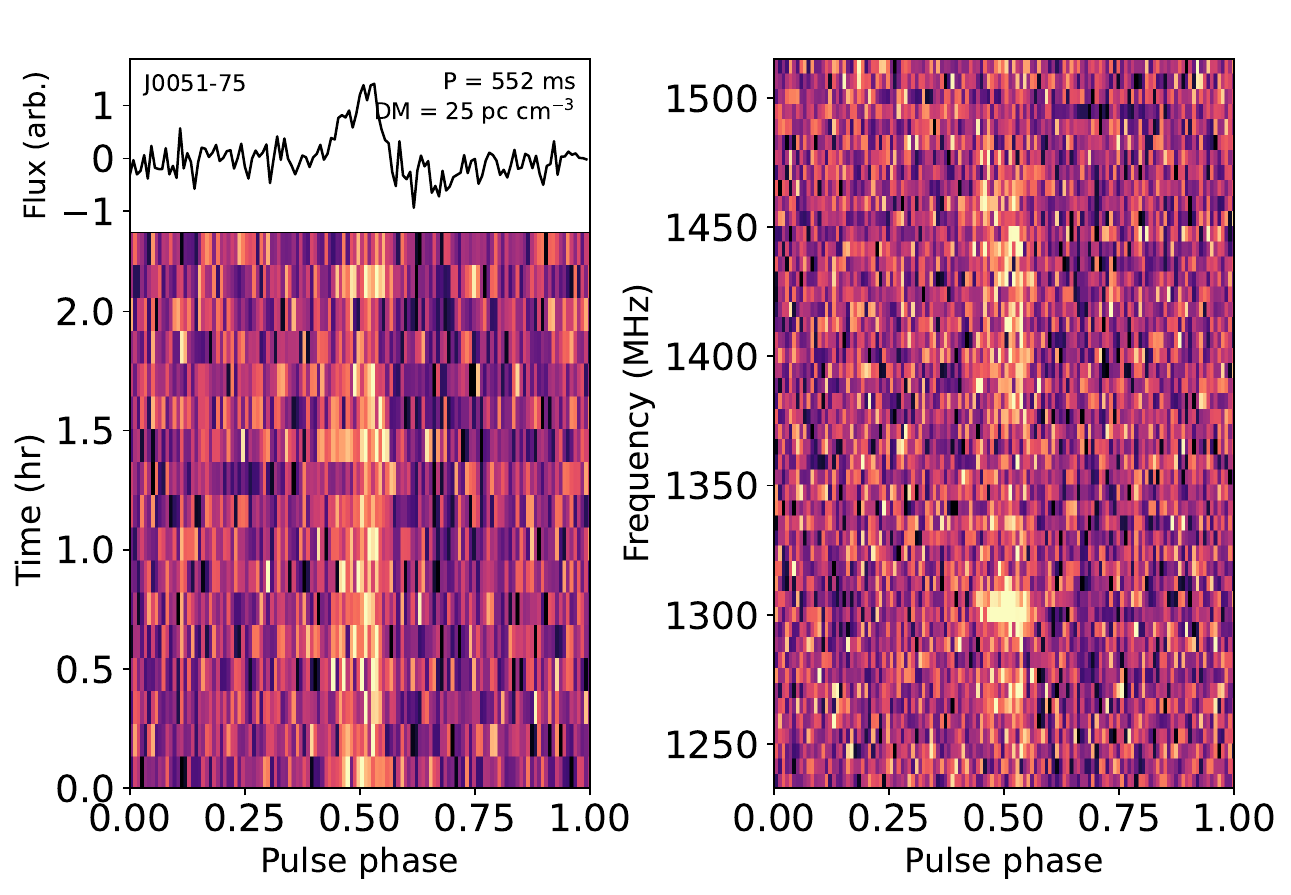}
\caption{Independent discovery plot of PSR~J0051$-$75 toward the SMC. The time--phase plot (left) shows 17 sub-integrations spanning 2.4\,hr, with noticeable variability in pulse brightness between sub-integrations. The frequency--phase plot (right) shows clear scintillation structure, consistent with the pulsar's relatively low DM of $25~{\rm pc~cm^{-3}}$.}
\label{fig:J0051_plot}
\end{figure}

\subsection{PSR~J0543--67}

PSR~J0543$-$67 was discovered with an $\mathrm{S/N}=13.6$, a spin period of $648~\mathrm{ms}$ and a DM of $70~\mathrm{pc\,cm^{-3}}$. As shown in the right panel of Figure~\ref{fig:discovery_plots}, emission is present only during the first $\sim 1.5~\mathrm{hr}$ of a $2.4~\mathrm{hr}$ integration, suggestive of nulling behavior. The signal also shows frequency-dependent modulation and is visible only up to $\sim1450~\mathrm{MHz}$, and there is no detectable emission in the uppermost $\sim70~\mathrm{MHz}$ of the band. This behavior is consistent with scintillation, which can cause variations in the observed pulsar intensity in both frequency and time \citep{rickett_90}. Therefore, the absence of the signal after $\sim1.5~\mathrm{hr}$ could also be caused by scintillation. To investigate whether this is the case, we removed the frequency range with no detectable emission and divided the remaining band into three sub-bands. In all three sub-bands, no signal is visibly detectable after $\sim1.5~\mathrm{hr}$, and the S/N values of these bands are consistent with non-detections ($\mathrm{S/N}<5$). The  absence of signal over different frequency ranges after 1.5\,hr is suggestive of broadband nulling \citep{gajjar_14, naidu_17}. However, given the limited S/N, we cannot distinguish conclusively between intrinsic nulling and scintillation. Additional detections with higher S/N are required to establish the origin of the observed time- and frequency-dependent behavior.
 
In order to confirm this pulsar, we conducted two $2.5~\mathrm{hr}$ follow-up observations with the Parkes UWL receiver, separated by one month, but did not detect the pulsar. The emission properties of this pulsar match those of PSR~J0443$-$68, which shows both nulling and scintillation in its discovery and confirmation observations. In addition, a known pulsar in the same dataset, PSR~J0451$-$67, shows similar band-limited emission (up to $1400~\mathrm{MHz}$) and was detected only twice over a two-year period \citep{manchester_06}. Given its relatively high discovery S/N, its time- and frequency-limited emission, and its consistency with the behavior of known nulling pulsars, we assert that the signal is indeed of astrophysical origin. Therefore, PSR~J0543$-$67 is likely to be a genuine pulsar despite the lack of follow-up confirmation detection.

\subsection{PSR~J0051--75}
PSR~J0051$-$75 is the only pulsar detected in the SMC data from the PM survey. The pulsar has a spin period of 552\,ms and a DM of $25~\rm pc\,cm^{-3}$. Although currently unpublished, the pulsar was first discovered in the SMART survey\footnote{\url{https://mwatelescope.atlassian.net/wiki/spaces/MP/pages/24970773/SMART+survey+candidates}}\citep{bhat_23} and designated as PSR~J0052$-$75. We therefore classify this source as an independent discovery. The pulsar has relatively low DM and shows clearly visible scintillation features (see Figure~\ref{fig:J0051_plot}). Since the pulsar was detected at relatively high significance (S/N = 22.4), we attempted to estimate its scintillation properties using the same method described in Section~\ref{subsec:J0443}. However, we could not reliably measure the scintillation bandwidth from the dynamic spectrum. The pulsar DM of $25~\rm pc\,cm^{-3}$ is well below the maximum Galactic DM predicted along this line of sight by the NE2001, YMW16, and NE2025 electron-density models ($\sim 45$, $\sim 33$, and $\sim 43~\rm pc\,cm^{-3}$, respectively), suggesting that the corresponding scattering predictions should be reasonably reliable in this case. For this line of sight, the predicted scintillation bandwidth is $\sim 3.7$\,MHz at 1\,GHz, which is comparable to our effective frequency resolution of $\sim 3$\,MHz. Therefore, any diffractive scintles would be only marginally resolved across the observing band, making a reliable measurement of the scintillation bandwidth difficult.

\begin{figure}
\centering
  \includegraphics[width=0.99\columnwidth]{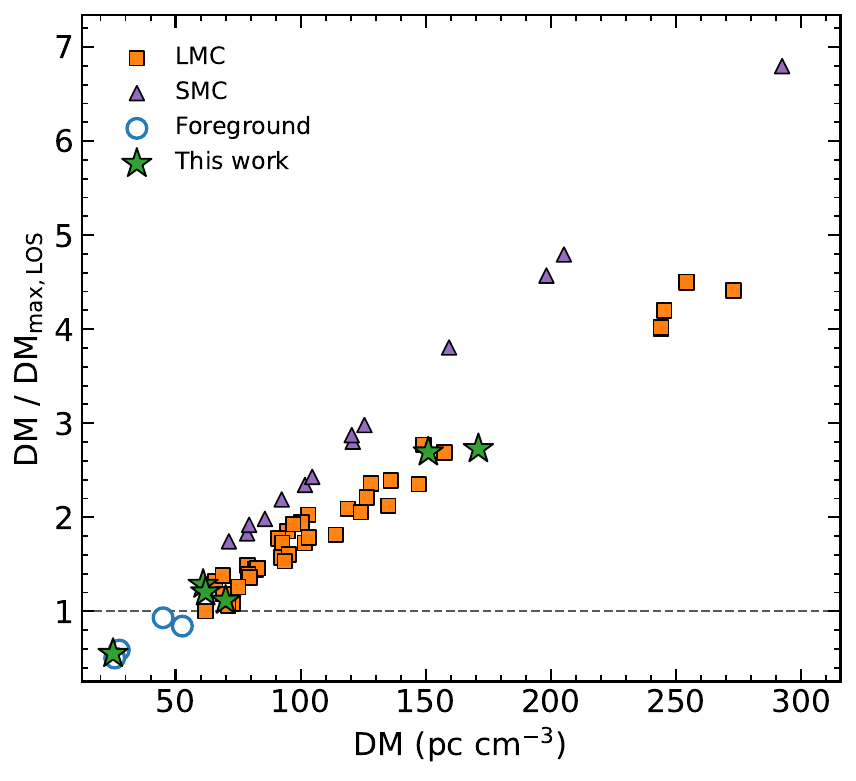}
\caption{Ratio of the DMs of the MC pulsars to the maximum Galactic line-of-sight DM, $\mathrm{DM}/\mathrm{DM}_{\rm max,LOS}$. For each pulsar, $\mathrm{DM}_{\rm max,LOS}$ is defined as the largest value predicted by the YMW16, NE2001, and NE2025 Galactic electron-density models.  The dashed line corresponds to the maximum DM expected from the Milky Way alone. The foreground pulsars lie below this threshold, whereas the LMC and SMC pulsars are above it.}
\label{fig:dm_ratio}
\end{figure}

\section{Discussion}
\label{sec:discussion}

\subsection{Known pulsar redetections}

In terms of known pulsar redetections, we identified a total of 24 pulsars in the PM survey data of the MCs. Of these, only five correspond to pulsars that had been reported in surveys conducted prior to \citet{manchester_06}. Fourteen pulsars were originally found in the first-pass processing of the PM survey by \citet{manchester_06}, and four additional pulsars were subsequently found in the reprocessing of the PM survey by \citet{ridley_13}. In addition, we redetected PSR~J0556$-$67, which was originally identified in the high-resolution survey data by \citet{hisano_22} at $\mathrm{RA} = 05^{\mathrm h}56^{\mathrm m}25.17^{\mathrm s}$, $\mathrm{Dec} = -67^\circ29'36.1''$ with a reported S/N of $\sim$29. However, in the PM data we also redetected this pulsar at $\mathrm{RA} = 05^{\mathrm h}54^{\mathrm m}49.51^{\mathrm s}$, $\mathrm{Dec} = -67^\circ40'36.4''$, with a lower S/N of $\simeq 11.3$. The angular separation between these two positions is $\simeq 14'$, which explains the reduced S/N in the PM survey. Overall, in this reprocessing we redetected all pulsars which were previously found in this survey.

\subsection{Association of the new pulsars with the LMC}

Of the six new pulsars, the observed DMs of five LMC pulsars are larger than the maximum Galactic line-of-sight DM ($\mathrm{DM}_{\rm max,LOS}$) contribution predicted by each of the three electron-density models, NE2001, YMW16, and NE2025 \citep[see Figure~\ref{fig:dm_ratio};][]{ne2001, ymw16, ocker_25}. Using the largest $\mathrm{DM}_{\rm max,LOS}$ returned by the three models, we find that $\mathrm{DM}/\mathrm{DM}_{\rm max,LOS}$ for these pulsars is between 1.1 and 2.7. The case is especially strong for PSRs J0520$-$6747 and J0541$-$6829, whose DMs are nearly a factor of three greater than $\mathrm{DM}_{\rm max,LOS}$. However, the DMs of PSRs J0443$-$6801, J0506$-$6657, and J0543$-$6737 lie closer to $\mathrm{DM}_{\rm max,LOS}$, but are systematically above the maximum foreground values predicted by all three models. Therefore, taken together, the DM excesses and their sky locations strongly suggest that all five new pulsars are associated with the LMC. For the remaining sixth pulsar, PSR~J0051$-$75, the DM is lower than $\mathrm{DM}{\rm max,LOS}$, with $\mathrm{DM}/\mathrm{DM}_{\rm max,LOS} \approx 0.5$, indicating that it is a Galactic foreground pulsar rather than being associated with the SMC.

\subsection{Flux densities and Luminosities}
\label{subsec:flux_density}

Since we do not yet have phase-connected timing positions or interferometric localizations, we estimated lower limits on the flux densities of the five pulsars in the LMC from their discovery pulse profiles, assuming the discovery position to be at the center of the beam. We used the radiometer equation \citep{dewey_85}, in which the estimated, or minimum detectable flux density (if position is unceratin) for a given detection is given by 

\begin{equation}
S_{\rm min} \;=\; 
\frac{(S/N)\,\beta\,(T_{\rm rec} + T_{\rm sky})}
     {G\,\sqrt{n_{\rm pol}\,\Delta\nu\,t_{\rm int}}}
\;
\sqrt{\frac{\delta}{1- \delta}} \, ,
\end{equation}

where the receiver temperature is typically  $T_{\rm rec} \approx 23$ K at 1.4\,GHz, and the total system temperature is $T_{\rm sys} = T_{\rm rec} + T_{\rm sky}$. Here, $T_{\rm sky}$ depends on the Galactic background and is generally a few Kelvin at high Galactic latitudes but can be significantly higher toward the Galactic plane. For the LMC we used a constant $T_{\rm sky}$ of 4.1\,K which is consistent with Global Sky Model estimates of $\sim 4$\,K at 1.4\,GHz toward the LMC and with values adopted in previous sensitivity calculations for Parkes LMC pulsar surveys. The telescope gain is $G =0.74$\,K\,Jy$^{-1}$, the observing bandwidth is $\Delta\nu \approx 288$\,MHz, the integration time per pointing is $t_{\rm int} = 8400$\,s, and $n_{\rm pol}=2$ corresponds to the two summed polarizations. The digitization factor $\beta$ accounts for losses due to finite-bit sampling and backend processing and is typically $\beta \approx 1.1$ for 2-bit data and 1.25 for 1-bit data. The duty cycle is defined as $\delta = W/P$, where $P$ is the pulse period and $W$ is the effective pulse width. The term $\left(\delta/(1-\delta)\right)^{1/2}$  reflects the reduction in sensitivity for broader pulses.

We find that PSR J0051$-$75, which is a foreground pulsar toward the SMC, has a flux density of $\sim150\,\mu$Jy. Its maximum DM-derived distance is estimated to be $\sim2.3$\,kpc, corresponding to a pseudo-luminosity of $\sim0.8\,\mathrm{mJy\,kpc^2}$. In contrast, all five LMC pulsars have relatively low flux densities, ranging from $\approx 26$--$60\,\mu$Jy. At an assumed LMC distance of $\sim$50\,kpc, these correspond to luminosities of $L_{1400}\sim 65$--150\,mJy\,kpc$^{2}$, and are close to the median of the known LMC pulsar population, which spans $\sim 30$ to $\sim 1000$\,mJy\,kpc$^{2}$ \citep[e.g.,][]{manchester_06, ridley_13, prayag_24}. A comparison of the 1.4-GHz flux densities of LMC pulsars from different surveys also validates this. The LMC pulsars discovered in the original Parkes survey of \citet{manchester_06} have a mean flux density of \(S_{1400} \approx 149\,\mu\mathrm{Jy}\). Following the reprocessing of those data by \citet{ridley_13}, four additional pulsars were identified, with a lower mean flux density of \(S_{1400} \approx 82\,\mu\mathrm{Jy}\). The five LMC pulsars reported here are even fainter, with a mean flux density of only \(S_{1400} \approx 38\,\mu\mathrm{Jy}\). For comparison, the currently known MeerKAT LMC pulsar sample has an even lower mean flux density of \(S_{1400} \approx 11\,\mu\mathrm{Jy}\) \citep{prayag_24, prayag_26}. Taken together, these values show a clear progression toward lower flux densities with increasingly sensitive searches and improved reprocessing techniques, as expected. In this context, the faintness of our new discoveries suggests that our reanalysis has been highly effective in uncovering pulsars that were previously too weak to be detected in the same dataset. 

Previous studies have compared the luminosity functions of the MC pulsars with those of the Galaxy and found apparent discrepancies. \citet{ridley_13} fitted the high-luminosity tail of the cumulative luminosity distribution with a power law, where the number of pulsars brighter than luminosity ($L$), $N(>L)$, can be expressed as
\[
\log N(>L) = \alpha \log L + C,
\]
and reported slopes of $\alpha = -1.2$ for Galactic pulsars with $L > 30$\,mJy\,kpc$^2$ and $\alpha = -3.6$ for LMC pulsars. The latter result, however, was based on a small LMC sample, and they emphasized that the difference was likely due to small-number statistics. More recently, \citet{wang_22} found that the slopes differ by about 23\% but are still statistically consistent. 

\begin{figure}
\centering
  \includegraphics[width=0.95\columnwidth]{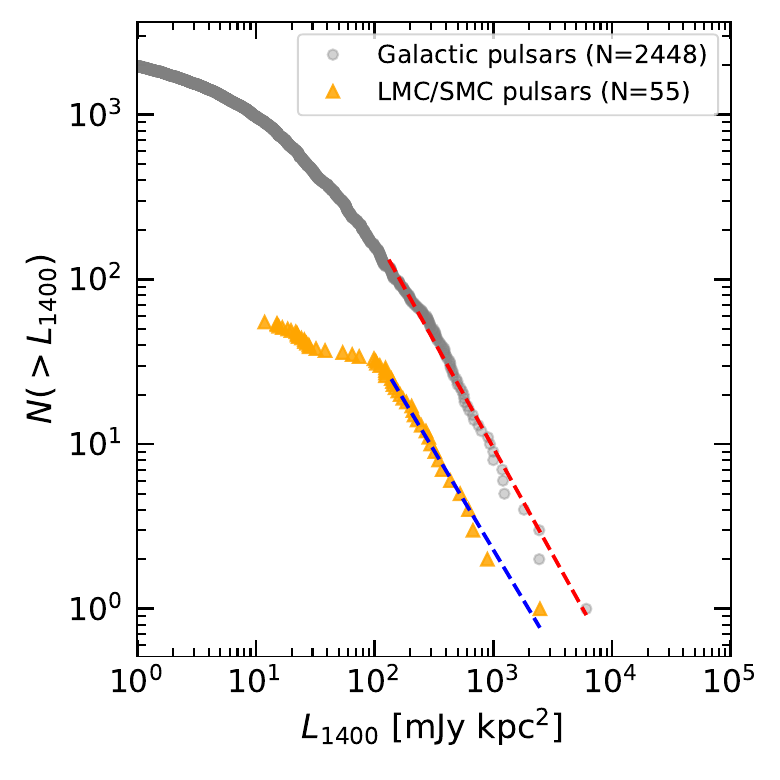}
  \caption{Cumulative distributions of radio luminosity for Galactic pulsars (gray circles) and MC pulsars (orange triangles) at 1400\,MHz. The ordinate shows the number of pulsars with luminosities greater than $L_{1400}$. The dashed lines represent power-law fits to the high-luminosity tails ($L_{1400} > 125$\,mJy\,kpc$^{2}$ for Galactic pulsars), with slopes of $\alpha=-1.30\pm0.01$ for the Galactic pulsars ($N=2448$) and $\alpha=-1.20\pm0.02$ for the MC pulsars ($N=55$). While all data points are shown, only those above the luminosity threshold were included in the fits.}
    \label{fig:clf_luminosity}
\end{figure}

Using the current enlarged sample of the MC pulsars (especially LMC pulsars), we re-examined the 1400\,MHz radio luminosity function by fitting the bright tail of the cumulative distributions with the same power-law model. For the Galactic sample, we excluded pulsars in globular clusters and those with \(P < 50\)\,ms, which resulted in 2{,}448 pulsars. For the MCs, we used 49 LMC pulsars (including the five reported in this work) and 7 in the SMC, resulting in a total of 56 pulsars for which flux densities—and hence luminosities—could be derived. To be consistent, for both populations we set the same luminosity threshold of \(125\,\mathrm{mJy\,kpc^{2}}\). For the Galactic pulsars we obtained a slope of \(\alpha = -1.30 \pm 0.01\), which is in good agreement with the slope obtained by \citet{ridley_13} and \citet{wang_22}. For the MC sample, we obtained a slope of \(\alpha = -1.20 \pm 0.02\), which is consistent with the Galactic luminosity function (see Figure~\ref{fig:clf_luminosity}). This implies that the luminosity functions of the Galactic pulsars and the combined MC pulsar population are nearly identical. However, when we treat the LMC and SMC populations separately, we obtain \(\alpha = -1.13 \pm 0.07\) for the LMC pulsars and \(\alpha = -0.9 \pm 0.1\) for the SMC pulsars. The slope for the LMC is statistically consistent with the Galactic value. Similarly, the SMC slope differs by about 30\% but remains consistent within the uncertainties. These results suggest that the luminosity functions of all three populations are broadly similar. The apparent differences are arguably due to small-number statistics in the MCs (particularly in the SMC) and distance uncertainties in the Galactic sample. With a growing number of MC pulsars, future studies will provide a more complete picture of their luminosity distribution.

\subsection{Duty Cycle–Period Relation}

In recent years, with the advent of sensitive surveys using MeerKAT and other reprocessing efforts including this work, the number of pulsars known in the MCs has nearly doubled. This provides an opportunity to revisit whether the pulsar population in the MCs differs from that of the Galaxy. As a first step, we examine the MC pulsars in terms of their pulse profile shapes. \citet{johnston_21} compared the pulse widths of these pulsars with those from the High Time Resolution Universe (HTRU) survey with the Parkes telescope \citep{keith_10}, which has a comparable sensitivity limit. They found that MC pulsar profiles appeared narrower than expected from the canonical Galactic period--width relation. However, they argued that this discrepancy likely arises from survey selection biases---particularly the preferential detection of narrower duty-cycle pulses---rather than from intrinsic differences in emission geometry. After accounting for these biases, the MC and Galactic profile-width distributions appear broadly consistent.

\begin{figure}
\centering
  \includegraphics[width=0.99\columnwidth]{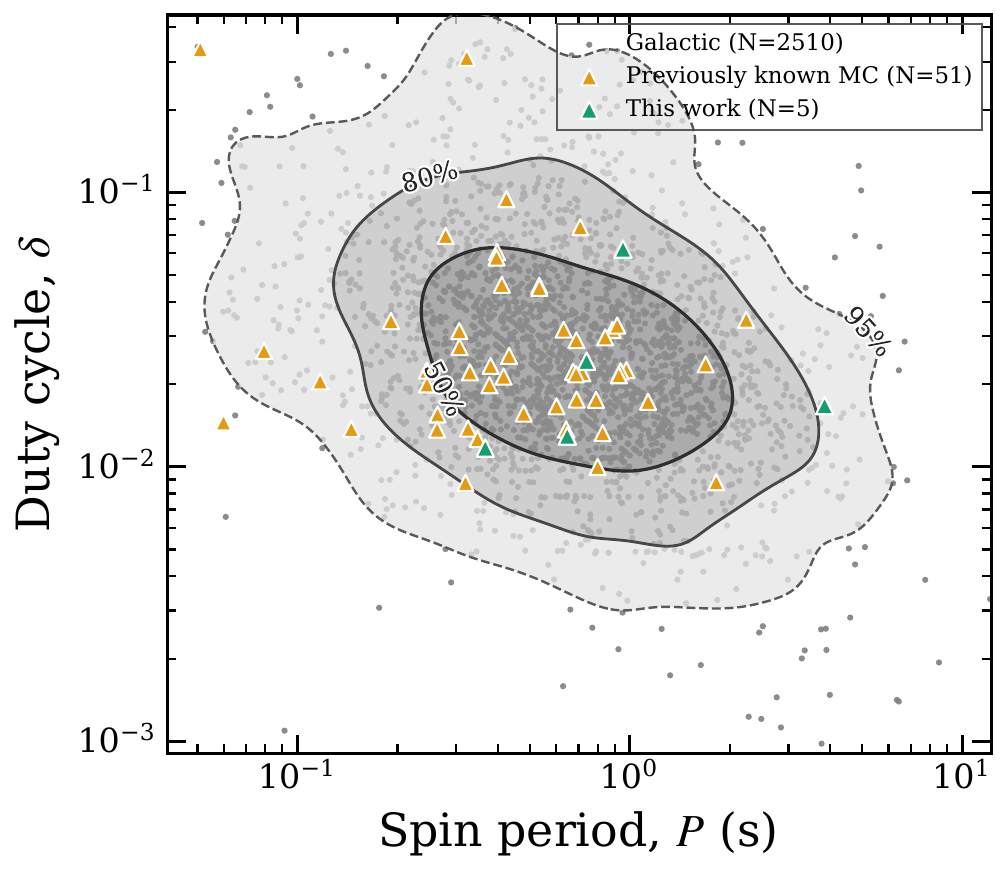}
  \caption{Duty cycle, $\delta$ as a function of spin period for Galactic pulsars (gray circles), previously known Magellanic Cloud pulsars (orange triangles), and the five pulsars reported in this work (green triangles). The shaded regions and contours show the 50\%, 80\%, and 95\% highest-density regions of a two-dimensional Gaussian kernel-density estimate fitted to the Galactic population in $(\log_{10} P, \log_{10} \delta)$ space.}
  \label{fig:dc_vs_period}
\end{figure}

With the growing sample of MC pulsars, we performed a similar test, but using the duty cycle--period relation. The Galactic pulsar sample was taken from the \textsc{PSRCAT}\footnote{\url{https://www.atnf.csiro.au/research/pulsar/psrcat/}} \citep{manchester_05}, excluding globular cluster pulsars and those with spin periods below 50\,ms. Duty cycles for the MC pulsars were also taken from \textsc{PSRCAT} where available. For five pulsars originally reported by \citet{ridley_13}, pulse widths were obtained from their redetections in our reprocessing.

In total, 56 MC pulsars were included in this analysis. Figure~\ref{fig:dc_vs_period} shows that the majority of the MC pulsars overlap with the Galactic pulsars and have duty cycles between 1--10\%. We used a non-parametric approach and constructed a two-dimensional Gaussian kernel-density estimate of the Galactic pulsar distribution in $(\log_{10}P,\log_{10}\delta)$ space, where $\delta=W_{50}/P$. For the observed MC sample, we calculated the summed log density
\[
S_{\rm MC} =
\sum_{i=1}^{N_{\rm MC}}
\ln f_{\rm Gal}(P_i,\delta_i),
\]
where $f_{\rm Gal}$ is the Galactic kernel density estimation (KDE). We then generated 20,000 independent samples of 56 pulsars from the Galactic KDE and calculated the same statistic for each realization. We find that the observed MC sample has a summed log likelihood of --35.17, compared with a median value of $-49.90$ for the simulated samples. The higher value of summed log likelihood for the MC sample indicates that the MC pulsars are more concentrated in the high-density regions of the Galactic period--duty-cycle distribution than a typical random sample drawn from the Galactic KDE. The observed value exceeds 96.9\% of the simulated values, while remaining within the 95\% range of the simulations (see Figure~\ref{fig:dc_vs_period}). This suggests that the observed MC pulsars are broadly similar to the Galactic pulsar population in the duty-cycle--period plane, and there is no significant evidence for a different distribution. 


Also, the concentration of the detected MC pulsars within the central part of the Galactic distribution is most likely due to observational selection effects. At the distances of the MCs, current pulsar surveys would detect the most radio-luminous pulsars, while pulsars with narrower duty cycles are generally easier to detect at a given mean flux density. Prior to the discovery of the 3.8\,s pulsar reported in this work, no MC pulsar with a spin period longer than 2.23\,s was known. The observed MC population is therefore likely affected by survey selection effects.

\subsection{Population synthesis study of the MCs}

\begin{table}
\centering
\caption{Population-synthesis predictions for the number of pulsars detectable in the MCs from the PM survey \citep{manchester_06} and from a future full survey with MeerKAT.}
\label{tab:population_synthesis_yields}
\begin{tabular}{llcc}
\hline
Galaxy & Survey & Normal pulsars & MSPs \\
\hline
SMC & Parkes multibeam & $3 \pm 1$ & $0$ \\
SMC & MeerKAT    & $35 \pm 6$ & $3 \pm 2$ \\
LMC & Parkes multibeam & $25 \pm 5$ & $0$ \\
LMC & MeerKAT   & $201 \pm 13$ & $5 \pm 2$ \\
\hline
\end{tabular}
\tablecomments{The MSP numbers should be regarded as upper limits.} The quoted uncertainties represent the standard deviation of the expected number of detections across the accepted simulation realizations.
\end{table}

Since the number of known radio pulsars in the MCs is now substantially larger than in earlier surveys, we revisited whether the PM survey has detected most of the pulsars expected at its sensitivity, and how many pulsars could be detected if a full survey of the Clouds were conducted with MeerKAT. Estimating the total pulsar population of the Clouds is not trivial because the star-formation histories of the LMC and SMC differ from that of the Milky Way. Therefore, we use a simpler, empirical population-synthesis approach using \texttt{PsrPopPy} \citep{bates_14} to predict the numbers of normal pulsars and MSPs detectable with the PM survey and with MeerKAT at L-band, since the current MeerKAT surveys only covered a small fraction of the LMC and SMC at coherent beam sensitivity. This is a snapshot calculation in which pulsars are assigned present-day radio properties rather than being evolved from birth in $P$--$\dot{P}$ space. We also make an empirical assumption that the underlying present day MC radio pulsar populations can be explained by the same radio property distributions as Galactic pulsars, effectively placing them at the MCs distances.

To simulate normal pulsars, we adopted the standard \texttt{PsrPopPy} period distribution with a mean log period of $2.7$ ms and a standard deviation of $0.34$, and its default log-normal 1400-MHz pseudo-luminosity prescription of \citet{faucher_giguere_06}, with a mean log luminosity of $-1.1$ and a standard deviation of $0.9$. This choice is motivated by our analysis in Section \ref{subsec:flux_density}, which shows that the high-luminosity distributions of the MC and Galactic normal-pulsar populations are broadly similar. We therefore retain this Galactic luminosity prescription rather than adopting the alternative model of \citet{szary_14} which can also be applied in a snapshot framework. However, its additional radio-efficiency criterion, $L_{\rm r}/\dot{E}<1\%$, requires $\dot{E}$, and hence $\dot{P}$, for each simulated pulsar. Our empirical snapshot population assigns $P$ but does not assign $\dot{P}$ or $\dot{E}$, and implementing the full prescription would therefore require introducing an additional present-day $\dot{P}$ or $\dot{E}$ distribution. The choice of luminosity distribution nevertheless represents an important systematic uncertainty. The \citet{szary_14} distribution, $\log_{10}L_{1400}\sim\mathcal{N}(0.5,1.0)$, is substantially brighter than the $\mathcal{N}(-1.1,0.9)$ distribution we used here and would place a larger fraction of pulsars in the high-luminosity tail detectable at MC distances, although the $<1\%$ radio-efficiency criterion would partly counteract this effect for low-$\dot{E}$ pulsars. Therefore, pulsar predictions should be interpreted with caution, as they are sensitive to the assumed luminosity distribution. For the simulated pulsars, we used a characteristic duty cycle of 3\%, in \texttt{PsrPopPy}, this is implemented through the pulse-width relation $W=0.03P^{0.9}$ with a log-normal scatter of roughly a factor of two variation in pulse width among individual pulsars. Spectral indices were drawn from a Gaussian distribution with a mean of $-1.41$ and a standard deviation of $0.96$ \citep{bates_13}. \citet{jankowski_18} found a somewhat steeper and narrower spectral-index distribution, although the effect of this choice is expected to be relatively small here because all surveys considered operate at similar L-band frequencies.

For MSPs, periods were drawn from a log-normal distribution centered near 4~ms and restricted to the range 1.5--30~ms. Since MSPs generally have broader pulse profiles, we used a Gaussian duty-cycle distribution with a mean of 12\% and a standard deviation of 4\%. Since no MSP has yet been detected in any MCs, their luminosity distribution is not observationally constrained. We therefore considered two different 1400-MHz luminosity prescriptions. We assumed that MSPs follow the same log-normal luminosity distribution as normal Galactic pulsars, and we also considered the power-law luminosity function of \citet{levin_13}. However, given very different star-formation histories and globular-cluster populations of MCs \citep[e.g.,][]{harris_04,harris_09,mackey_04}, our adopted power law assumptions for MSPs could be highly uncertain and can be significantly different from the Milky Way.

Sky positions were drawn from two-dimensional Gaussian distributions centered on the respective Clouds and were used to determine whether a pulsar lies within the survey footprint and its angular offset from the nearest survey pointing. Since the MCs, and therefore the pulsars within them, are effectively at similar distances compared with Galactic pulsars, we assigned distances from Gaussian distributions centered at 50~kpc for the LMC and 62~kpc for the SMC. To allow for the finite depth of the Clouds along the line of sight, we used distance dispersions of 2.5~kpc for the LMC and 4~kpc for the SMC. These values are consistent with the few-kpc depth measured for the LMC and the larger, approximately 4~kpc depth of the SMC \citep{hashcke_12, subramanian_09, deb_09}. Since no detailed electron-density model exists for the MCs, we adopted a simple DM distribution centered at $\sim$120~pc~cm$^{-3}$ with a dispersion of 40~pc~cm$^{-3}$. Scattering is included through the \citet{bhat_04} relation implemented in \texttt{PsrPopPy}.

Using the above assumptions, we first tested whether the simulations could reproduce the number of pulsars detected by the PM survey in both Clouds. For the LMC, \citet{ridley_13} estimated a population of $57\,000^{+70\,000}_{-30\,000}$ radio-emitting normal pulsars beaming towards Earth (95\% credibility), so we adopted a rounded trial population of 50\,000, which lies within this range. For the SMC, we adopted a parent population of 16,000 normal pulsars, consistent with the estimate of \citet{titus_20}.

For comparison with the supernova rates of the Clouds, adopting an approximate radio-active lifetime of 50--100~Myr, these normal-pulsar populations beaming towards Earth correspond to formation rates of $\sim0.05$--$0.10$ pulsars per century in the LMC and $\sim0.016$--$0.032$ pulsars per century in the SMC.
These rates are below the estimated SN/SNR rates of $\sim0.2$--$0.5$ per century in the LMC \citep{leahy_17,bozzetto_17} and $\sim0.08$ per century in the SMC \citep{leahy_22}. However, correcting for pulsars whose beams do not intersect our line of sight and for pulsars lost through natal kicks would increase the inferred total birth rates, and could lead to tension with the SN rate \citep{ridley_10}. A full comparison would additionally depend on the fraction of core-collapse events producing neutron stars rather than black holes and the fraction of neutron stars \citep{heger_03}. For MSPs we adopted an upper limit parent population of 5500 in the LMC and 8000 in the SMC \citep{ridley_10}, therefore, the predicted MSP detection should also be regarded as MSP detection upper limits. Also these numbers correspond to pulsars that remain within the clouds. An estimated 46\% and 67\% of LMC and SMC born pulsars may have been ejected by natal kicks and we have not included them \citep{ridley_10}.

We first passed each realization through the survey available before the PM survey, when only five LMC pulsars were known. Only realizations that reproduced approximately this number of detections were retained, after which the PM survey parameters were applied to the same accepted population. The same procedure was used for the SMC using the pulsars found before the PM survey as the initial constraint. For each cloud, 100 accepted realizations were generated, and each realization was repeatedly attempted, up to the adopted maximum number of trials, until a population satisfying the calibration criterion was obtained. In total, this required 392 trial populations for the LMC and 207 for the SMC, of which 100 were retained in each case, corresponding to acceptance fractions of 25.5\% and 48.3\%, respectively. This predicts $25\pm5$ normal pulsars in the LMC and $3\pm1$ in the SMC for the PM survey. These predictions are in good agreement with the observed pulsars presented in this work, where including both the redetections and new pulsars, we redetected a total of 21 pulsars in the LMC and 5 pulsars in the SMC. This shows that the adopted population assumptions and survey-selection model can reproduce the observed PM pulsar population.

We then used the PM survey as the calibration survey for the MeerKAT predictions. Simulation realizations were retained only when the PM survey model reproduced the observed pulsar population in each Cloud, and the same accepted realization was subsequently passed through the MeerKAT survey model. The same underlying parent populations of normal pulsars and MSPs were used for both the PM and MeerKAT calculations, with no additional scaling of the population for MeerKAT.

For the PM survey simulations, we used the observing parameters and survey footprint of \citet{manchester_06} which covered most of the main bodies of the MCs. \texttt{PsrPopPy} accounts for the angular offset of each simulated pulsar from the nearest survey pointing and reduces its S/N according to the telescope beam response. For the MeerKAT survey, we used TRAPUM-like L-band parameters following \citet{prayag_24} for the LMC and \citet{carli_24} for the SMC, with integrations of two hours per pointing. Since, no pulsar in the MCs has ever been discovered below a folded S/N of 10, therefore, for all survey configurations, we adopted a detection threshold of ${\rm S/N}=10$. This choice is also supported by our reprocessing in which several candidates with \({\rm S/N}\sim9\) were followed up but could not be confirmed as pulsars. Since the exact coherent-beam tiling depends on the array configuration and beam layout, we approximated its effect through the effective survey area and assumed that 75\% of the adopted regions in both Clouds are covered by the coherent-beam tiling. This is consistent with the $\sim78\%$ coverage above half-power obtained for a 50\% beam-overlap configuration in the MMGPS simulations \citep{padmanabh_23}. Therefore, the MeerKAT predictions should be interpreted as the expected yields for the adopted survey regions under TRAPUM-like sensitivity and our coverage assumptions.

The simulation results using the Galactic log-normal luminosity prescription for both normal pulsars and MSPs are summarized in Table~\ref{tab:population_synthesis_yields}. For the SMC, a comparable MeerKAT survey is expected to detect $35 \pm 6$ normal pulsars. For the LMC, where 24 pulsars (including the new discoveries reported in this work) were known prior to the MeerKAT targeted searches in MCs, the MeerKAT simulations predict a much larger detectable population of $201\pm13$ normal pulsars. However, for MSPs the predicted number strongly depends on the assumed luminosity function. If we assume MSPs in MCs follow the same log-normal luminosity distribution as normal Galactic pulsars, we find  $3\pm2$ detections in the SMC and $5\pm2$ in the LMC. In contrast, using the Galactic MSP power-law luminosity function of \citet{levin_13} reduces the expected yield to essentially zero in the SMC and only about 0.5 MSPs in the LMC.

In summary, these population-synthesis results suggest that the PM survey has likely recovered most of the normal pulsars accessible to its sensitivity, although the underlying pulsar population in the MCs is far from exhausted. A full MeerKAT survey with TRAPUM-like sensitivity could uncover a substantially larger pulsar population, particularly in the LMC, and may also enable the first detection of an MSP in the Clouds.

\section{Conclusions}
\label{sec:conclusion}

We have reprocessed the archival PM survey \citep{manchester_06} as well as the high-time-resolution survey of the MCs \citep{ridley_13}. Although these data have been reprocessed previously, our FFT-based search techniques, combined with effective candidate filtering and inspection, have yielded six additional pulsars. Five are located in the LMC and one is a foreground pulsar toward the SMC. This increases the known LMC pulsar population to a total of 49. These discoveries demonstrate the continued potential of reprocessing legacy survey data to uncover scientifically valuable pulsars when searched with improved pipelines and candidate-selection techniques. In particular, our FFT-based pipeline pushes sensitivity to faint signals through deep harmonic summing (up to the 32nd harmonic) and a candidate-selection stage that preserves coverage in period, spectral S/N, and harmonics prior to final folding.

Although nulling and long-term intermittency are well established within the broader Galactic pulsar population, pulsars in the MCs have more commonly been characterized by strong flux-density variability attributable to scintillation \citep{johnston_21}. Previously, nulling in MC pulsars had been reported only for PSR~J0529$-$6652 \citep[e.g.,][]{crawford_13}, while PSR~J0451$-$67 appears to be highly intermittent as it has been detected only twice over two years of timing observations \citep[e.g.,][]{manchester_06}. More recently, PSR~J0540$-$69 has also been shown highly variable and apparently intermittent emission, being detected in only a small fraction of archival observations covering its position and exhibiting substantial flux-density variations on timescales of tens of minutes \citep{xia_25}. In this context, PSR~J0443$-$68 reported in this work also shows clear evidence of both nulling and scintillation. This pulsar was initially identified only as a faint candidate with an estimated flux density of $\sim26~\mu$Jy. In a follow-up confirmation observation, it was detected with $\sim75~\mu$Jy and exhibited a null lasting $\sim36$ min (see Figure~\ref{fig:J0443-68_plots}).

PSR~J0506$-$66 shows time- and frequency-dependent emission variability, including a $\sim30$ min interval of undetectable emission in the discovery observation. The possible frequency-dependent variability could be caused by scintillation, while the interval of undetectable emission may indicate nulling. However, given its low S/N, it is difficult to distinguish intrinsic nulling from a scintillation-induced low-flux state. PSR~J0543$-$67 also shows possible time- and frequency-dependent emission variability. In this case, the signal becomes undetectable after $\sim1.5$\,hr over the same time interval in three sub-bands, which is suggestive of broadband nulling, although scintillation cannot be completely excluded at the current S/N. The pulsar also remained undetected in two subsequent $2.5$\,hr Parkes observations, and additional observations will be required to determine whether these non-detections are caused by intrinsic intermittency, scintillation, or a combination of both.

Together, these sources add to the growing evidence that nulling and intermittent emission, which are commonly observed in Galactic pulsars, are also present among pulsars in the MCs. However, because the low-flux states of PSRs~J0506$-$66 and J0543$-$67 cannot yet be unambiguously attributed to intrinsic nulling rather than scintillation, the current sample is insufficient to derive a reliable nulling fraction for the LMC pulsar population. These discoveries also provide a natural explanation for why such sources can be missed in earlier processings and can fail to be redetected in follow-up observations. At extragalactic distances, frequency- and time-dependent variability can move a weak pulsar signal above or below the detection threshold of a finite-duration observation, while intrinsic intermittency or long null states, where present, can further reduce the probability of detection.

Using an updated LMC pulsar sample that incorporates our new discoveries together with recent MeerKAT/TRAPUM detections, we revisited the 1.4-GHz luminosity distribution. Within current uncertainties, the LMC luminosity function is consistent with that of the Galactic non-recycled pulsar population: the fitted power-law slopes agree at the $\sim 15$\% level, and the apparent steepening reported in \citet{ridley_13} and \citet{wang_22} can be attributed to small-number statistics and heterogeneous survey depths. We also obtain a similar conclusion from the duty-cycle--period comparison where on average MC pulsars appear slightly narrower at a given period, the difference is plausibly explained by selection effects (distance-limited detection of the brightest, narrowest sources) rather than by a fundamentally distinct emission geometry in the MCs.


Finally, the population synthesis results suggest that the PM survey has detected most of the normal pulsars accessible to its sensitivity. At the same time, the simulations indicate that a significant underlying pulsar population remains to be discovered in both the LMC and SMC. Our simulations further show that a full MeerKAT survey at L-band would increase the MC pulsar population substantially, with predicted yields of $201 \pm 13$ normal pulsars in the LMC and $35 \pm 6$ in the SMC. Such a survey may also lead to the first detections of MSPs in the MCs, with expected yields of $5 \pm 2$ and $3 \pm 2$ MSPs in the LMC and SMC, respectively. However, the predicted MSP yields should be regarded as upper limits. Detecting MSPs in these galaxies would be particularly exciting, as they could provide useful additions to pulsar timing arrays by extending them beyond the Milky Way, which will provide additional long-baseline sight lines for future low-frequency gravitational-wave investigations. Furthermore, they could be useful probes of the interstellar and intergalactic medium and could enable measurements of electron density distributions, magnetic field structures, and propagation effects through both the Galactic halo and the MCs themselves, while also preparing the groundwork for future extragalactic pulsar studies with next-generation telescopes.

\begin{acknowledgments}
We thank the anonymous referee for their constructive comments and suggestions, which helped improve the quality of this manuscript. We thank David Kaplan for the useful comments. R.S.\ is thankful for the continuing support from the Max Planck Society. L.V. acknowledges support from the National Science Foundation (NSF) Physics Frontiers Center award numbers 1430284 and 2020265. MB was supported by ARC grant CE230100016, the OzGrav ARC Centre of Excellence for gravitational wave discovery. The Parkes radio telescope is part of the Australia Telescope National Facility, funded by the Australian Government for operation as a National Facility managed by CSIRO. We acknowledge the Wiradjuri People as the Traditional Owners of the Parkes Observatory site. The Australia Telescope Compact Array is also part of the Australia Telescope National Facility (\url{https://ror.org/05qajvd42}), and we acknowledge the Gomeroi People as the Traditional Owners of its site. 
\end{acknowledgments}

\vspace{5mm}
\facilities{Parkes/Murriyang radio telescope}

\software{astropy \citep{2013A&A...558A..33A,2018AJ....156..123A}, PSRCHIVE \citep{hotan_2004}, PEASOUP, RIPTIDE, PRESTO, Pulsar Survey Scraper \citep{2022ascl.soft10001K}, PyGEDM \citep{2021PASA...38...38P}.}

\bibliography{lmc}{}
\bibliographystyle{aasjournal}

\end{document}